\documentclass[aps,prc,reprint,showkeys,superscriptaddress,twocolumn]{revtex4}

\usepackage{graphicx}
\usepackage{enumitem}
\usepackage{epstopdf}
\usepackage{multirow}
\usepackage{booktabs}
\usepackage{verbatim}

\begin{document}


\title{The modified astrophysical S-factor of the ${}^{12}$C+${}^{12}$C fusion reaction at sub-barrier energies}


\author{Y. J. Li}
\affiliation{China Institute of Atomic Energy, Beijing 102413, China}

\author{X. Fang}
\email[]{fangx26@mail.sysu.edu.cn}
\affiliation{Sino-French Institute of Nuclear Engineering and Technology, Sun Yat-sen University, Zhuhai, Guangdong 519082, China}

\author{B. Bucher}
\affiliation{Idaho National Laboratory, Idaho Falls, ID 83415, USA}

\author{K. A. Li}
\author{L. H. Ru}
\author{X. D. Tang}
\affiliation{Institute of Modern Physics, Chinese Academy of Sciences, Lanzhou, Gansu 730000, China}
\affiliation{School of Nuclear Science and Technology, University of Chinese Academy of Sciences, Beijing 100049, China}



\begin{abstract}
The $^{12}$C+$^{12}$C fusion reaction plays a crucial role in stellar evolution and explosions. Its open reaction channels mainly include $\alpha$, $p$, $n$, and ${}^{8}$Be. Despite more than a half century of efforts, large discrepancies remain among the experimental data measured using various techniques. In this work, we analyze the existing data using the statistical model. Our calculation shows: 1) the relative systematic uncertainties of the predicted branching ratios get smaller as the predicted ratios increase; 2) the total modified astrophysical S-factors (S$^*$ factors) of the $p$ and $\alpha$ channels can each be obtained by summing the S$^*$ factors of their corresponding ground-state transitions and the characteristic $\gamma$ rays while taking into account the contributions of the missing channels to the latter. After applying corrections based on branching ratios predicted by the statistical model, an agreement is achieved among the different data sets at $\emph{E}_{\rm cm}>$4 MeV, while some discrepancies remain at lower energies suggesting the need for better measurements in the near future. We find that the recent S$^*$ factor obtained from an indirect measurement is inconsistent with the direct measurement at energies below 2.6 MeV. We recommend upper and lower limits for the ${}^{12}$C+${}^{12}$C S$^*$ factor based on the existing models. A new $^{12}$C+$^{12}$C reaction rate is also recommended. 

\end{abstract}

\keywords{carbon burning, $^{12}$C+$^{12}$C fusion, statistical model, quasi-molecular resonance, astrophysical S-factor, reaction rate, stellar evolution}

\maketitle


\section{Introduction}
\label{sec_introduction}
The ${}^{12}$C+${}^{12}$C reaction at astrophysical energies is a crucial reaction for stellar evolution and explosion. For example, stars with mass more than 8 solar masses can ignite the ${}^{12}$C+${}^{12}$C fusion reaction and proceed with carbon burning inside their cores when the core temperature reaches above 0.6 GK. These stars end their lives as Ne/O white dwarfs. More massive stars will continue the ${}^{12}$C+${}^{12}$C reaction in their shells at temperatures around 1.0-1.2 GK and eventually become supernovae. When a star explodes, a shock wave propagates through the outer layers of the dying star and initiates the explosive carbon burning which imprints its unique nucleosynthetic patterns in the ashes of the dying star. The ${}^{12}$C+${}^{12}$C fusion reaction is considered to be the ignition reaction of type Ia supernovae \cite{mori2018} and superbursts. In type Ia supernovae, ignition happens in the white dwarf core typically at T$\sim$ 0.15-0.7 GK and $\rho\sim$ (2-5)$\times$10$^9$ g/cm$^3$. In type-I X-ray bursts, ash from the rp-process builds up on the surface of the neutron star with a significant amount of ${}^{12}$C (3$\sim$10\%). Heat sources in the crust of neutron star raise the temperature of the ash and eventually trigger carbon ignition at a temperature of $\sim$0.5 GK~\cite{cooper2009} and density above 3$\times$10$^9$ g/cm$^3$. The ignition conditions mentioned above strongly depend on the actual ${}^{12}$C+${}^{12}$C reaction rate as well as the estimation of the screening effect in dense matter\cite{gasques2005,gasques2007,cooper2009}.

\begin{figure}[htbp]
\centering
\includegraphics[width=0.49\textwidth]{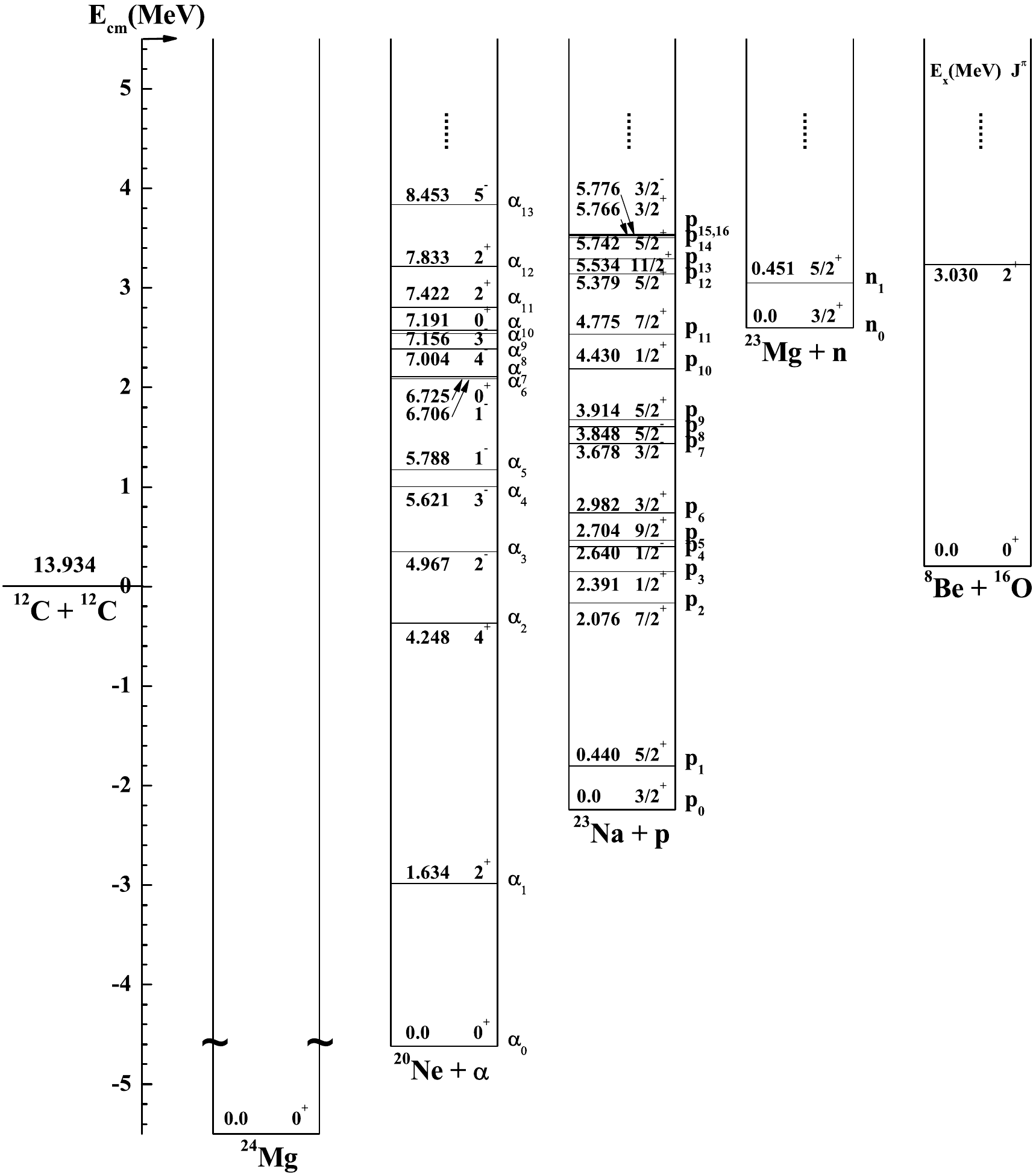}
\caption{Energy-level diagram for the $^{12}$C+$^{12}$C system with the primary exit channels at low energies. The p$_{i}$, $\alpha_{i}$, and n$_{i}$ represent the protons produced with ${}^{23}$Na, $\alpha$ particles produced with ${}^{20}$Ne, and neutrons produced with ${}^{23}$Mg, respectively, at the ground state ($i$=0) or the $i^{th}$ excited state($i$=1,2,3,...). The energies for most common characteristic $\gamma$ rays are 440 keV for $^{23}$Na, 1634 keV for ${}^{20}$Ne, and 450 keV for ${}^{23}$Mg. Other than these three most important channels, the ${}^{12}$C(${}^{12}$C,$^8$Be)${}^{16}$O channel is also possible and may warrant further investigation.}
\label{12C12C_energy_level}
\end{figure}

The crucial energy range extends from a few tens keV to $\emph{E}_{\rm cm}$=3 MeV \cite{gasques2007}, well below the Coulomb barrier at $\emph{E}_{\rm cm}$=5.5 MeV \cite{notani2012}. Two ${}^{12}$C nuclei fuse into the compound nuclear states of ${}^{24}$Mg with excitation energies of 14 to 17 MeV. The compound states then decay through five channels:
\begin{eqnarray*}
{}^{12}\rm C+{}^{12}\rm C &\rightarrow& {}^{20}{\rm Ne}+\alpha+4.62\text{ MeV} \nonumber
\\                          &\rightarrow& {}^{23}{\rm Na}+{\rm p}+2.24\text{ MeV} \nonumber
\\			   &\rightarrow& {}^{23}\rm Mg+{\rm n}-2.60\text{ MeV} \nonumber
\\                          &\rightarrow& {}^{16}{\rm O}+{}^{8}{\rm Be}-0.20\text{ MeV}
\\                          &\rightarrow& {}^{24}{\rm Mg}+14.934\text{ MeV}
\end{eqnarray*}
The related energy levels of the compound nucleus and residual nuclei are shown in Fig.~\ref{12C12C_energy_level}. The energies of protons and alphas in the decay channels are above their Coulomb barriers when the excitation energy of the compound nucleus is above the $^{12}$C+${}^{12}$C separation energy. As a result, the particle decay widths are much larger than the $\gamma$-decay width. Therefore, the contribution of the radiative capture channel is negligible.

%
${}^{12}$C(${}^{12}$C,$\alpha$)${}^{20}$Ne and ${}^{12}$C(${}^{12}$C,$p$)${}^{23}$Na are the two major reaction channels at sub-barrier energies\cite{patterson1969,sandorfi1978,nathan1981}. The measurements of ${}^{12}$C(${}^{12}$C,$\alpha$)${}^{20}$Ne and ${}^{12}$C(${}^{12}$C,$p$)${}^{23}$Na at sub-barrier energies can be classified into two categories: the detection of either characteristic $\gamma$-rays or light charged particles. A summary of previous experimental work directly measuring $\alpha$- and $p$-channels of the $^{12}$C+$^{12}$C reaction near stellar energies is listed in Table \ref{tab_summary}. A pioneering particle spectroscopy experiment was done by Patterson $\emph{et al.}$ \cite{patterson1969} who measured the cross sections of $p$ and $\alpha$ using a telescope system consisting of a proportional counter and a silicon detector. Later Mazarakis and Stephens \cite{mazarakis1973}, and Becker $\emph{et al.}$ \cite{becker1981} separately repeated the particle spectroscopy experiments using silicon detectors. Limited by the target purity and beam induced backgrounds, these measurements were held to the range $\emph{E}_{\rm cm}>$2.7 MeV. Using a thick highly ordered pyrolytic graphite (HOPG) \cite{novoselov2005,soldano2010} target, Zickefoose $\emph{et al.}$ \cite{zickefoose2018} successfully suppressed these backgrounds and extended the particle-spectroscopy measurement of ${}^{12}$C(${}^{12}$C,p$_{0,1}$)${}^{23}$Na down to $\emph{E}_{\rm cm}$=2.0 MeV. No other channels were measured due to a thick degrader used in front of the detectors to control the beam induced background.
For the particle detection experiment, the protons and $\alpha$ particles corresponding to highly excited fusion residues are emitted with relatively small kinetic energies in the laboratory frame and are often ignored because their energies are below detection thresholds or they are overwhelmed by large backgrounds at lower energies. Therefore, it is necessary to evaluate the contribution from this missing part of the fusion cross section measurement. 

Detection of characteristic $\gamma$-rays emitted by the fusion residues is also an effective way to identify their production. In the ${}^{12}$C+${}^{12}$C fusion reaction, the most common characteristic $\gamma$ rays are 440 keV for $^{23}$Na, 1634 keV for ${}^{20}$Ne. Such measurements have been done by Kettner $\emph{et al.}$ \cite{kettner1980} and Aguilera $\emph{et al.}$ \cite{aguilera2006} using HPGe in the range of $\emph{E}_{\rm cm}>$2.6 MeV. The main limitations are backgrounds from cosmic rays and reactions with target impurities. By using a plastic veto detector and removing the target impurities with a high intensity beam current, Spillane $\emph{et al.}$ \cite{spillane2007} extended the $^{12}$C+$^{12}$C fusion down to $\emph{E}_{\rm cm}$=2.1 MeV and reported a very strong resonance at $\emph{E}_{\rm cm}$=2.14 MeV. To further suppress the cosmic-ray background and some beam-induced background, Jiang $\emph{et al.}$ \cite{jiang2018} developed the particle-$\gamma$ coincidence experiment using a silicon array and $\gamma$-array. Most recently, two newly measured results have been reported by Fruet $\emph{et al.}$ \cite{fruet2020} and Tan $\emph{et al.}$ \cite{tan2020} using the particle-$\gamma$ coincidence technique.

Unlike the particle detection experiment, the $\gamma$-spectroscopy experiment is not a complete measurement of the total fusion cross section. The ground states ($p_0$, $\alpha_0$, $n_0$) do not emit any $\gamma$ rays, while some excited states decay with significant branching through transitions that bypass the main characteristic $\gamma$ rays. Spillane $\emph{et al.}$ estimated the contributions of the missing channels using results from the particle spectroscopic measurements. Aguilera $\emph{et al.}$ \cite{aguilera2006} suggested an approach in which the total cross sections for the $^{12}$C( $^{12}$C,p)$^{23}$Na and $^{12}$C($^{12}$C,$\alpha$)$^{20}$Ne channels are assumed to be the summation of the cross sections of the 440 keV and 1634 keV $\gamma$ rays and the cross sections of the ground states ($p_0$ and $\alpha_0$), respectively. Another $\gamma$-ray technique was used by Dasmahapatra $\emph{et al.}$ \cite{dasmahapatra1982} to measure the total fusion cross section. First they measured the partial cross sections using a $\gamma$ summing detector. Then this partial cross section was converted into the total fusion cross section with the aid of the statistical model. However the systematic errors of their approach have rarely been discussed. 

The probability of decay through the ${}^{12}$C(${}^{12}$C,$n$)${}^{23}$Mg channel is weaker than the $p$ and $\alpha$ channels because of its negative Q-value. This reaction is believed to play an important role in the carbon shell burning of massive stars \cite{woosley2002,pignatari2012}. The important energy range for astrophysics is 2.7$<\emph{E}_{\rm cm}<$3.6 MeV. The reaction was first studied by Patterson $\emph{et al.}$ \cite{patterson1969} who measured the cross section over the range $\emph{E}_{\rm cm}$=4.23 to 8.74 MeV by counting the $\beta$-rays from the ${}^{23}$Mg decays. Dayras $\emph{et al.}$ \cite{dayras1977} measured the cross sections down to $\emph{E}_{\rm cm}$=3.54 MeV by counting the $\gamma$-rays emitted following the ${}^{23}$Mg beta decay. Bucher $\emph{et al.}$ \cite{bucher2015,bucher2014} further extended the measurement down to $\emph{E}_{\rm cm}$=3.0 MeV, deep within the Gamow window, by counting the neutrons. They also developed a theoretical prediction of the cross sections at lower energies with a systematic uncertainty of about 40\% \cite{bucher2015}. These experimental and theoretical data have the necessary precision required to reliably model various hydrostatic and explosive carbon shell burning scenarios.

The ${}^{12}$C(${}^{12}$C,$^8$Be)${}^{16}$O (or ${}^{12}$C(${}^{12}$C,2$\alpha$)${}^{16}$O) channel is a difficult one to study. The characteristic $\gamma$-ray method does not work because the excitation energy of the first excited state of ${}^{16}$O ($\emph{E}_{\rm x}$=6.13 MeV) is too high to be populated at sub-barrier energies. The $\alpha$ particles from the disintegration of ${}^8$Be have energies down to nearly zero. Therefore, it is extremely difficult to study this channel. The only experiment was done by {\v{C}}ujec $\emph{et al.}$ \cite{cujec1989} who measured the low energy $\alpha$-particles at forward angles using a tracking foil in the range of $\emph{E}_{\rm cm}$=2.425 to 5.24 MeV. Their results showed that at $\emph{E}_{\rm cm}<$3.13 MeV, the ${}^8$Be becomes larger than the $^{12}$C($^{12}$C,$\alpha_0$)$^{20}$Ne cross section, suggesting that $\alpha$-transfer is favored at such low energies. However, this result has never been confirmed by others. Another possibility to this channel is a two-step process via  ${}^{12}$C(${}^{12}$C,$\alpha$)${}^{20}$Ne. When the excitation energy of $^{20}$Ne exceeds 5.621 MeV, the $\alpha$ decay channel of $^{20}$Ne dominates. This 2$\alpha$ channel has been studied down to 3.2 MeV by detecting the first $\alpha$ in the two-step process~\cite{mazarakis1973}. Our statistical model calculation in this paper suggests the contribution from either of these channels is negligible at $\emph{E}_{\rm cm}<$3.2 MeV.



\begin{figure*}[htbp]
\centering
\includegraphics[width=1\textwidth]{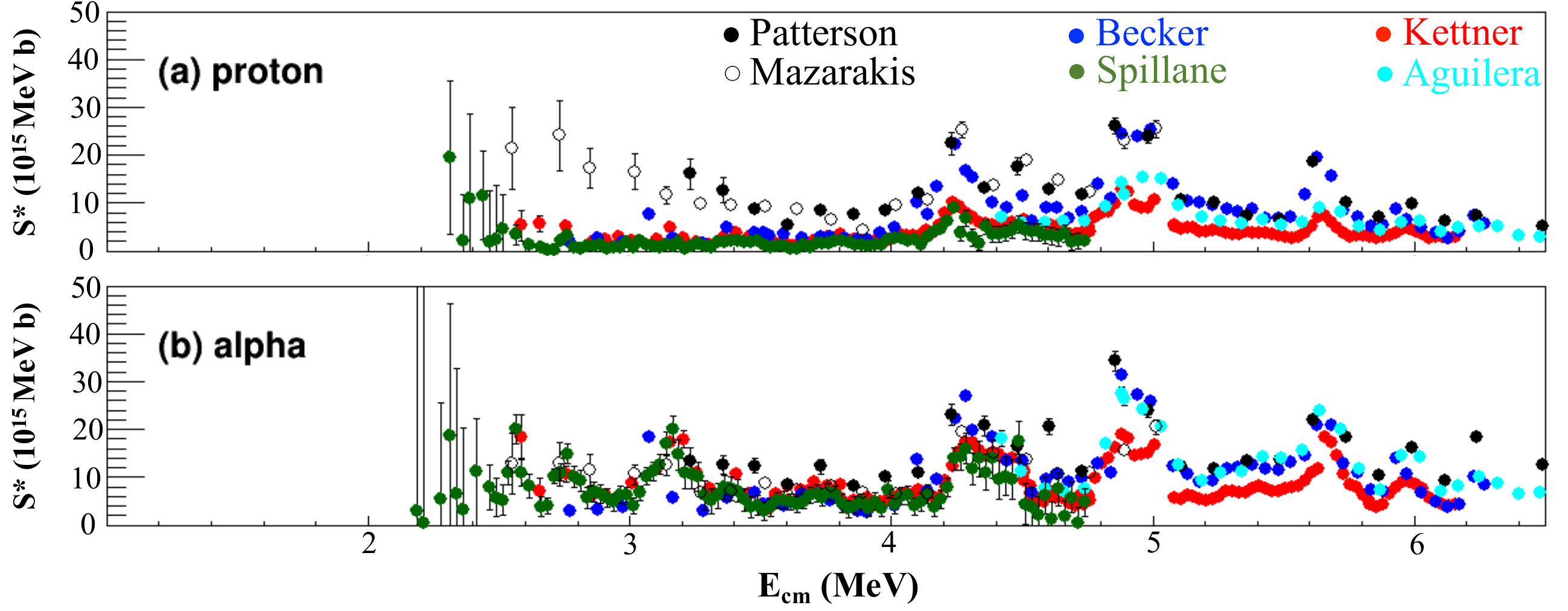}
\caption{S$^*$ factor of the measured cross sections for the $p$ and $\alpha$ channels, respectively. The S$^*$ factors measured by Kettner $\emph{et al.}$ (red solid circle), Aguilera $\emph{et al.}$ (cyan solid circle) and Spillane $\emph{et al.}$ (green solid circle) are based on the characteristic $\gamma$-ray technique. The S$^*$ factors measured by Becker $\emph{et al.}$ (blue solid circle), Patterson $\emph{et al.}$ (black solid circle) and Mazarakis $\emph{et al.}$ (black open circle) are based on the particle spectroscopy.}
\label{fig_sfactor_old}
\end{figure*}
\begin{figure*}[htbp]
\centering
\includegraphics[width=1\textwidth]{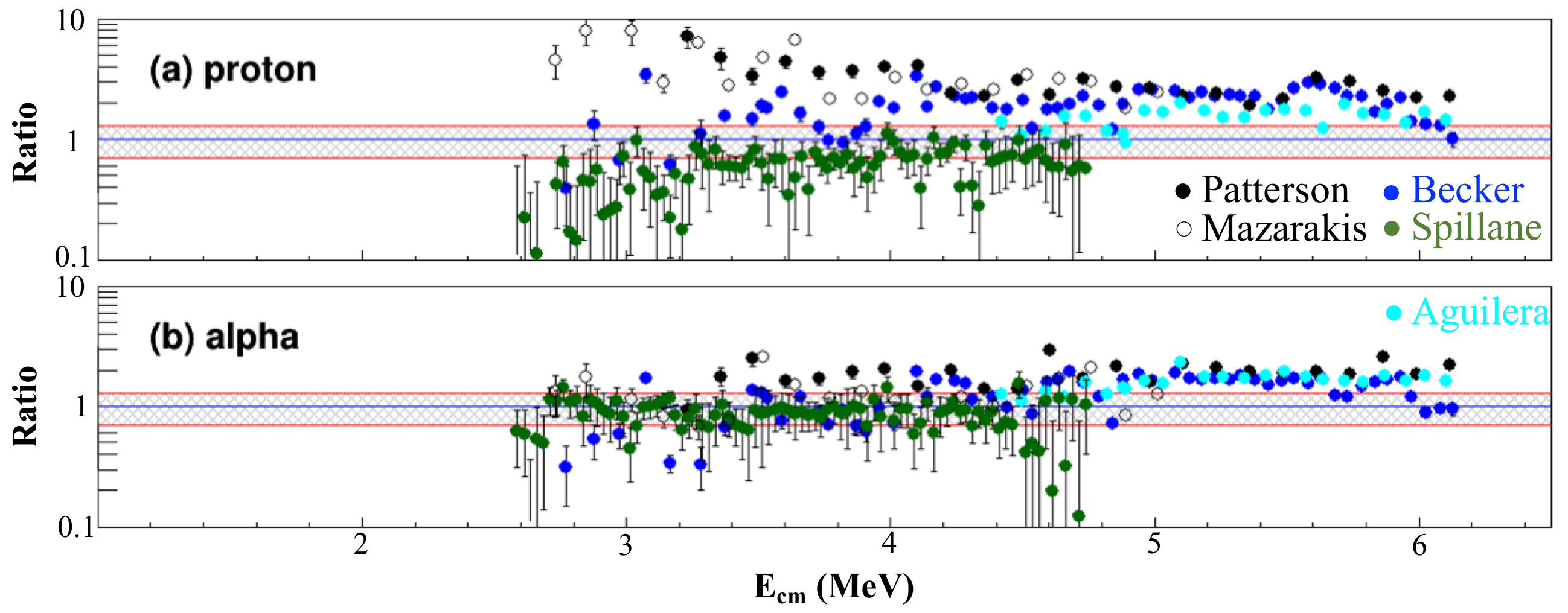}
\caption{Ratios of the various data sets as shown in Fig. \ref{fig_sfactor_old} to the baseline data set, which is the measurement by Kettner $\emph{et al.}$. The means and errors of the baseline data sets are interpolated in the calculations of ratios. The shaded areas shown in the ratio plots are corresponding to the deviation of $\pm$30\%.}
\label{fig_ratio_old}
\end{figure*}

We convert the cross sections of ${}^{12}$C(${}^{12}$C,$\alpha$)${}^{20}$Ne and ${}^{12}$C(${}^{12}$C,$p$)${}^{23}$Na into the S$^*$ factor \cite{patterson1969},
\begin{equation}
S^*(\emph{E}_{\rm cm})=\sigma(\emph{E}_{\rm cm}) \emph{E}_{\rm cm} exp( \frac{87.21}{\sqrt{\emph{E}_{\rm cm}}}+0.46 \emph{E}_{\rm cm}),
\end{equation}
to remove most of the Coulomb barrier penetration effect and show the complicated nuclear structure. The results are shown as Fig.\ref{fig_sfactor_old} and Fig.\ref{fig_ratio_old}, in which all data are the measured partial cross sections without branching ratio corrections. For the proton channel, it includes the partial cross section of the 440 keV characteristic $\gamma$-ray of $^{23}$Na and the sum of the cross sections of the detected proton channels. For the alpha channel, it includes the partial cross section of the 1634 keV characteristic $\gamma$-ray of $^{20}$Ne and the sum of the cross sections of the detected alpha channels. There is a 1636 keV $\gamma$-ray emitted by the ${}^{23}$Na. This contribution is included within the partial cross section of the 1634 keV $\gamma$-ray of ${}^{20}$Ne. The sums of the detected $p$/$\alpha$ cross sections only reflect the partial cross sections of the production of ${}^{23}$Na/${}^{20}$Ne. Some $p$/$\alpha$ are missing because their energies are too low to be detected. As a result, large discrepancies exist among the different data sets. For the proton channel, when $\emph{E}_{\rm cm}>$3.6 MeV, the S$^*$ factors measured with the particle spectroscopy by Patterson $\emph{et al.}$ and Becker $\emph{et al.}$ are about a factor of 2 higher than the S$^*$ factors measured from the characteristic $\gamma$-rays by Kettner $\emph{et al.}$. But the differences raise up to a factor of 4--9 at energies around 3 MeV. This observation demonstrates the importance of the $\gamma$-ray decay branching ratios. 

There are also differences existing among the S$^*$ factors obtained using the same experimental techniques. Taking the particle spectroscopy as an example, the summed S$^*$ factors of the proton channel of Becker $\emph{et al.}$, Patterson $\emph{et al.}$ and Mazarakis $\emph{et al.}$ agree with each other  within $\pm$20\% for $\emph{E}_{\rm cm}>$4 MeV. But the S$^*$ factors of Becker $\emph{et al.}$ are lower than the other two data sets by a factor of up to 10 when the energies are below 3.4 MeV. The difference at high energies may be explained by the normalization uncertainties, such as target thickness and/or beam intensity. The difference at lower energies, on the other hand, mainly comes from the fact that Becker $\emph{et al.}$ measured less channels than the other two groups. Aguilera $\emph{et al.}$, Spillane $\emph{et al.}$, Kettner $\emph{et al.}$ all used the characteristic $\gamma$-rays to measure the S$^*$ factors. Compared to the S$^*$ factors of Kettner $\emph{et al.}$, the S$^*$ factors of Agulera $\emph{et al.}$ are about 30\% higher while those of Spillane $\emph{et al.}$ are more than (30--60)\% lower. The only significant difference in the experimental procedure of these three works is that Aguilera $\emph{et al.}$ placed the Ge detector at 55$^{\circ}$ to minimize the effect of the $\gamma$-ray angular distribution, while the other two groups chose 0$^{\circ}$ as their detection angle. The difference between the results of Spillane $\emph{et al.}$ and Ketter $\emph{et al.}$ shows this effect of angular distribution is less than 30\%.

Measures have been taken in some experiments to account for the missing channels in order to obtain the actual total S$^*$ factor. For example, Becker $\emph{et al.}$ summed all the observed particle channels to obtain the total fusion cross sections. In cases where the data for a given particle group were only available over a limited energy range, the energy-averaged S$^*$ factors of these groups were extrapolated down to threshold and added to the total S$^*$ factors. 
Based on the result of Becker $\emph{et al.}$, Spillane $\emph{et al.}$ estimated the mean values of the ratios of the 440 keV line to the sum of all the proton channels  and the 1634 keV line to the sum of all the alpha channels to be 0.55$\pm$0.05 and 0.48$\pm$0.05, respectively. This correction is included in their reported S$^*$ factors of the proton and alpha channels. Aguilera $\emph{et al.}$ took a different approach by adding the S$^*$ factors of the $p_0$ and $\alpha_0$ channels to their measured S$^*$ factors of the 440 keV and 1634 keV $\gamma$-rays to obtain the total S$^*$ factors. Using their own data set as the standard, Aguilera $\emph{et al.}$ shifted the energies and adjusted the normalization factors to bring a reasonable agreement among the existing experimental results. 

In this paper, we introduce a new approach based on the statistical model to predict the branching ratio of each decay channel. The prediction is validated by the experimental data obtained by particle spectroscopy. The branching ratios are predicted for each experiment and used to convert the observed S$^*$ factors into the total S$^*$ factor of ${}^{12}$C+${}^{12}$C. 

\begin{table*}[htbp]
\caption{\small Summary of experimental measurements of $\alpha$- and p- channels in the ${}^{12}$C+${}^{12}$C fusion near stellar energies.}
\label{tab_summary}
\begin{center}
\resizebox{1\textwidth}{!}{
\begin{tabular}{p{2.3cm} |p{1.6cm} |p{1.6cm} |p{3cm}| p{4.5cm}| p{4.5cm}| p{4cm}}
\toprule[0.75pt] \noalign{\smallskip}
Measurement & $\emph{E}_{\rm cm}$(MeV) & Beam & Target & Technique &Detection of partial $\sigma$ & Method to calculate total $\sigma$ \\
\noalign{\smallskip}\hline\noalign{\smallskip}

Patterson \cite{patterson1969} & 3.23 -- 8.75  & 0.017--0.17 p$\mu$A & Carbon foil with thickness of about 40 $\mu$g/cm$^2$. & Particle spectroscopy. Protons and $\alpha$ particles were detected at four angles between 20$^{\circ}$ to 80$^{\circ}$ by a $\Delta$E-E telescope consisting of an argon proportional counter 6 cm long and a silicon surface-barrier detector 1.5 mm thick. The n-channel was measured by counting the offline $\beta$-decay of residual $^{23}$Mg above 4.25 MeV. & Total cross sections: $\sigma_{\alpha}$, $\sigma_{p}$, $\sigma_n$. & $\sigma_{\alpha}$=sum of $\alpha_0$, $\alpha_1$, $\alpha_2$, $\alpha_3$, $\alpha_4$, $\alpha_5$; $\sigma_{p}$=sum of $p_{0-1}$, $p_{2-6}$, $p_{7-9}$, $p_{10}$. \\  \hline

Mazarakis and Stephens \cite{mazarakis1973} & 2.55 -- 5.01  & $\leq$0.38 p$\mu$A & Self-supporting foils from high-purity graphite, 30, 53, 65 $\mu$g/cm$^2$.  &Particle spectroscopy. Surface barrier Si detectors at eight angles between 20$^{\circ}$ and 90$^{\circ}$. & Cross sections of $\alpha_0$, $\alpha_1$, $\alpha_2$, $\alpha_3$, $\alpha_4$+$\alpha_5$, $\alpha_6$, $\alpha_7$, and $p_0$, $p_1$, $p_2$, $p_3$, $p_4$+$p_5$, $p_6$, $p_7$, $p_8$+$p_9$, $p_{10}$. & $\sigma_{\alpha}$=$\sum \sigma_{\alpha_i}$, $\sigma_{p}$=$\sum \sigma_{p_i}$. \\ \hline

Becker \cite{becker1981} & 2.8 -- 6.3  & 1--5 p$\mu$A & 
Self-supporting foils from graphite, 8 to 30 $\mu$g/cm$^2$. & Particle spectroscopy. Nine surface barrier Si detectors at 10$^{\circ}$ to 90$^{\circ}$ (in 10$^{\circ}$ steps). & Cross sections of $\alpha_0$, $\alpha_1$, $\alpha_2$, $\alpha_3$, $\alpha_4$, $\alpha_5$, $\alpha_6$, $\alpha_7$, $\alpha_8$+$\alpha_9$, $\alpha_{10}$, $\alpha_{11}$, $\alpha_{12}$, and $p_0$, $p_1$, $p_2$, $p_3$, $p_4$+$p_5$, $p_6$, $p_7$, $p_8$+$p_9$, $p_{10}$, $p_{11}$, $p_{12}$, $p_{13}$, $p_{14}$+$p_{15}$+$p_{16}$.  & $\sigma_{\alpha}$=$\sum \sigma_{\alpha_i}$, $\sigma_{p}$=$\sum \sigma_{p_i}$. Estimating the contributions of the missing channels with the extrapolation of the averaged S$^*$ factor at higher energies  \\ \hline

Zickefoose \cite{zickefoose2018} & 2.0 -- 4.0  & $\leq$15 p$\mu$A & Thick target: high-purity graphite; HOPG. & Particle spectroscopy. Two $\Delta$E--E telescopes at 130$^{\circ}$ (􏰁$\Delta$E--Si detector: area A=300 mm$^2$, thickness t=15 $\mu$m; E--Si detector: A=300 mm$^2$, t=300 $\mu$m).  & The $p_0$+$p_1$ reaction yield of the infinitely thick target was obtained with energy steps
of $\Delta$􏰁E= 20 to 100 keV. & No total cross sections were given. \\ \hline \hline


Kettner \cite{kettner1980} & 2.45 -- 6.15 & $\leq$15 p$\mu$A  & Carbon targets (9 to 55 $\mu$g/cm$^2$) were evaporated on 0.3mm thick Ta backings. & $\gamma$􏰍-ray spectroscopy. One Ge(Li) detector at 0$^{\circ}$.  & $\gamma$-ray transitions from a large number of excited states in $^{20}$Ne, $^{23}$Na and $^{23}$Mg were observed, providing corresponding partial cross sections. &  \\ \hline

Dasmahapatra \cite{dasmahapatra1982} & 4.2 -- 7.0 & $\leq$0.015 p$\mu$A & Carbon foil with thickness of about 30 $\mu$g/cm$^2$.  &  $\gamma$􏰍-ray spectroscopy. Two NaI detectors, in almost 4$\pi$ geometry. The pulses from the two NaI detectors were summed via a amplifier. & The total-$\gamma$-ray-yield method. & To obtain total cross sections, using the fraction $\sigma_{\gamma}$/$\sigma_{total}$ from experimental data of Mazarakis and Stephens \cite{mazarakis1973}. \\ \hline


Aguilera \cite{aguilera2006} & 4.42 -- 6.48 &  & Amorphous C-foil deposited onto thick Ta backing. Thickness: 19.2$\pm$0.9 $\mu$g/cm$^2$, 22.7$\pm$1.0 $\mu$g/cm$^2$, 29.9$\pm$1.4 $\mu$g/cm$^2$. & $\gamma$􏰍-ray spectroscopy. Two HPGe at 125$^{\circ}$ and 55$^{\circ}$.  & $\gamma$-ray cross sections of 1634, 440, 450 keV, corresponding to ground-state transitions. & The $\alpha_0$ and $p_0$ cross section values of Becker $\emph{et al.}$ \cite{becker1981} were used to correct data. $\sigma_{\alpha}$=$\sigma_{\alpha_0}$+$\sigma_{\gamma(1634)}$, $\sigma_{p}$=$\sigma_{p_0}$+$\sigma_{\gamma(440)}$. \\ \hline

Spillane \cite{spillane2007} & 2.10 -- 4.75 & $\leq$40 p$\mu$A  & Thick target: high-purity graphite. & $\gamma$􏰍-ray spectroscopy. One HPGe at 0$^{\circ}$. The $\gamma$-ray efficiencies: (3.6$\pm$0.4)\% for $\emph{E}_{\gamma}$=440 keV, (1.9$\pm$0.2)\% for $\emph{E}_{\gamma}$=1634 keV.  & The $\gamma$􏰍-ray thick-target yields of the 440 and 1634 keV lines with energy steps of 􏰒12.5 to 25 keV. To arrive at a thin-target yield􏰉, the thick-target yield curve was differentiated. &  $\sigma_{\alpha}$= $\sigma_{\gamma(1634)}$, $\sigma_{p}$= $\sigma_{\gamma(440)}$.  \\ \hline \hline

Jiang \cite{jiang2018} & 2.68 -- 4.93 & $\leq$0.6 p$\mu$A &Isotopically enriched (􏰅99.9\%) $^{12}$C targets with thickness of about 30-50 $\mu$g/cm$^2$.  & Particle-$\gamma$ coincidence technique. DSSD$_{1,2,3}$: 147$^{\circ}$--170$^{\circ}$, 123$^{\circ}$--143$^{\circ}$, 17$^{\circ}$--32$^{\circ}$; 25\% of 4$\pi$. The $\gamma$-ray efficiencies: 9\% for $\emph{E}_{\gamma}$=440 and 7\% for $\emph{E}_{\gamma}$=1634 keV. & $\sigma_{p_1}$, $\sigma_{p_2}$, and incomplete $\sigma_{p_3}$, $\sigma_{p_4}$, $\sigma_{p_5}$, $\sigma_{p_6}$, $\sigma_{p_{7,8,9}}$; $\sigma_{\alpha_1}$, $\sigma_{\alpha_2}$, and incomplete $\sigma_{\alpha_3}$. & Normalized by ratios $\sigma_{total}$/$\sigma_{p_1}$, $\sigma_{total}$/$\sigma_{\alpha_1}$ \emph{etc.} from Becker $\emph{et al.}$ \cite{becker1981} and Mazarakis and Stephens \cite{mazarakis1973} \\ \hline

Fruet \cite{fruet2020} & 2.16, 2.54--3.77, 4.75--5.35 & $\leq$2 p$\mu$A & Carbon foil using rotating target mechanism, with thickness of about 20-70 $\mu$g/cm$^2$.  & Particle-$\gamma$ coincidence technique. Three annular silicon strip detectors covering 30\% of the 4$\pi$ solid angle. An array of 36 LaBr$_3$(Ce) scintillator detectors. & $\sigma_{p_1}$, $\sigma_{\alpha_1}$. & Normalized by ratios $\sigma_{p_1}$/$\sigma_{p}$=(15.6$\pm$0.7)\%, $\sigma_{\alpha_1}$/$\sigma_{\alpha}$=(31.9$\pm$1.4)\%. \\ \hline

Tan \cite{tan2020} & 2.2, 2.65--3.0, 4.1-- 5.0 & $\leq$13 p$\mu$A & Thick target: HOPG.  & Particle-$\gamma$ coincidence technique. six YY1s: 102$^{\circ}$--146$^{\circ}$; one S2: 151$^{\circ}$--170$^{\circ}$. The $\gamma$-ray efficiencies: 2.3\% for $\emph{E}_{\gamma}$=440 and 1.22\% for $\emph{E}_{\gamma}$=1634 keV. & $\sigma_{p_1}$, $\sigma_{\alpha_1}$. & Normalized by ratios $\sigma_{p_1}$/$\sigma_{p}$, $\sigma_{\alpha_1}$/$\sigma_{\alpha}$ from Becker $\emph{et al.}$ \cite{becker1981}. \\
\noalign{\smallskip} \bottomrule[0.75pt]
\end{tabular}
}
\end{center}
\end{table*}

\section{The statistical model calculation}
\label{sec_HF}

\begin{figure}[htbp]
\centering
\includegraphics[width=0.45\textwidth]{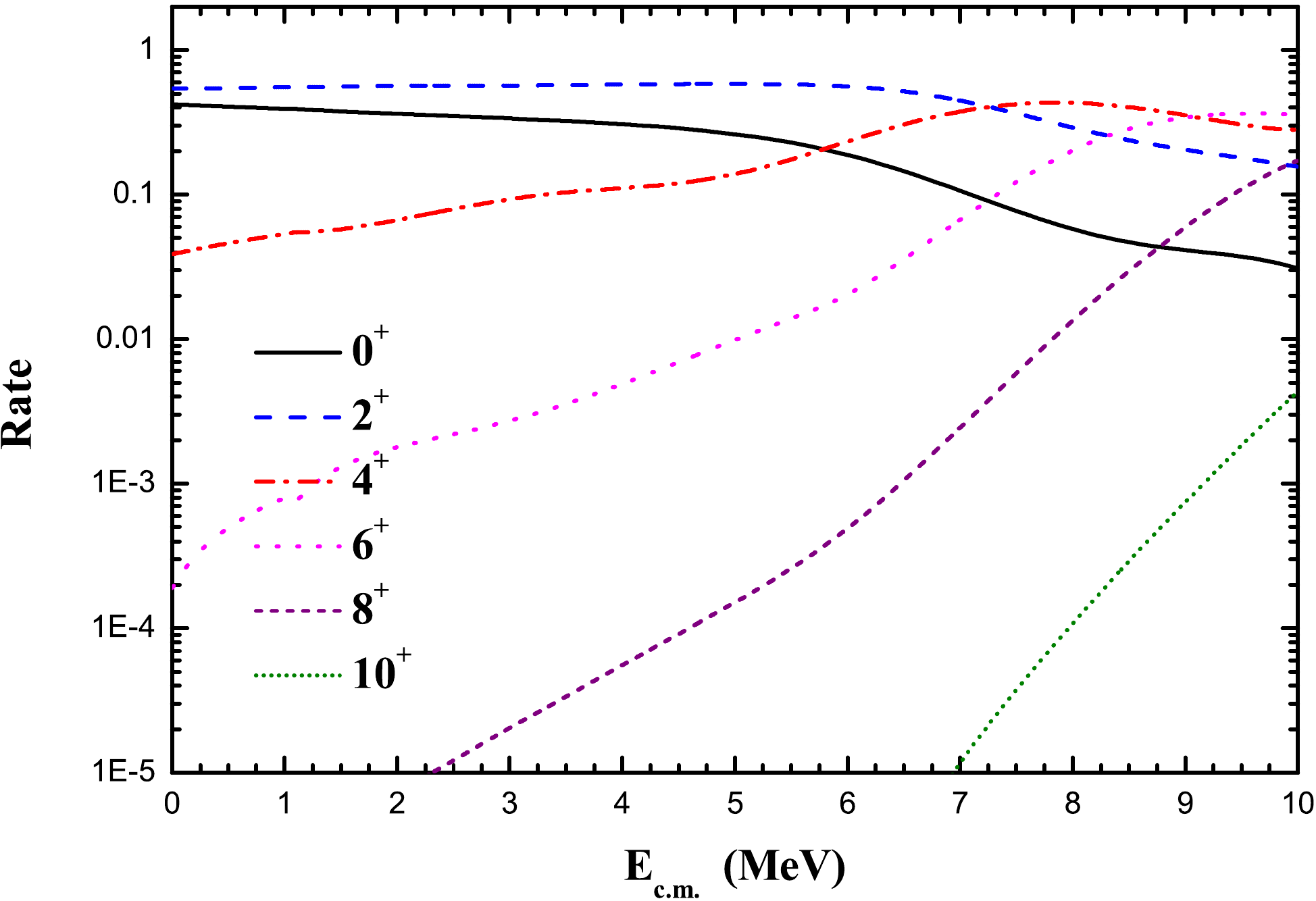}
\caption{Spin populations of the $^{24}$Mg compound nucleus made by the ${}^{12}$C+${}^{12}$C fusion, calculated with the CCFull code \cite{hagino1999}. The dominant components have spins $0^+$, $2^+$, $4^+$ and $6^+$.}
\label{spin_population_24Mg}
\end{figure}

The spin populations of the $^{24}$Mg compound nucleus are calculated by fitting the average S$^*$ factor of ${}^{12}$C+${}^{12}$C with a simple powered Woods-Saxon potential. The fusion cross sections are calculated by using the CCFull code \cite{hagino1999}. The corresponding spin populations of the $^{24}$Mg compound nucleus are shown in Fig. \ref{spin_population_24Mg}. Only even spins with positive parity are allowed since the ${}^{12}$C+${}^{12}$C system is composed of two identical bosons. The calculation shows that the spin of the compound nucleus is dominated by the 2$^+$ and 0$^+$ states at energies below $\emph{E}_{\rm cm}$=3 MeV. Theoretical calculations show that some spins could be enhanced at certain energies by the molecular resonances in the entrance channel. The smooth spin population shown here, however only represents the average behavior. 


Let all quantum numbers that specify the colliding nuclei and the two nuclei in the exit channel be denoted by

The fusion evaporation cross sections for different decay channels are modeled with the Hauser-Feshbach formula \cite{hauser1952,vogt1968}. Let all quantum numbers that specify the entrance and exit channels are denoted by $\alpha$ and $\alpha^\prime$, respectively. Similarly, \textbf{l} + \textbf{S} = \textbf{J} = \textbf{l$^\prime$} + \textbf{S$^\prime$}, \textbf{S} = \textbf{I} + \textbf{i}, and \textbf{S$^\prime$} = \textbf{I$^\prime$} + \textbf{i$^\prime$} denote the angular momentum coupling for orbital angular momentum $l$ ($l^\prime$), channel spin $S$ ($S^\prime$), total angular momentum $J$, and intrinsic angular momenta $I$ ($I^\prime$) and $i$ ($i^\prime$) \cite{stokstad1981}. The Hauser-Feshbach formula is expressed as

\begin{equation}
\sigma_{\alpha\alpha'}=\pi\lambda\hspace{-.6em}\bar{~~}_\alpha^2\sum_J\frac{2J+1}{(2I+1)(2i+1)}\frac{[\Sigma_{Sl}T_l(\alpha)]^J[\Sigma_{S'l'}T_l(\alpha')]^J}{[\Sigma_{a'',S''l''}T_{l''}(\alpha'')]^J}
\label{Hauser_Feshbach_formula}
\end{equation}

\noindent for the angle integrated cross section, where $T_l$ denotes the optical model transmission coefficient.

\begin{figure*}[htbp]
\centering
\includegraphics[width=0.9\textwidth]{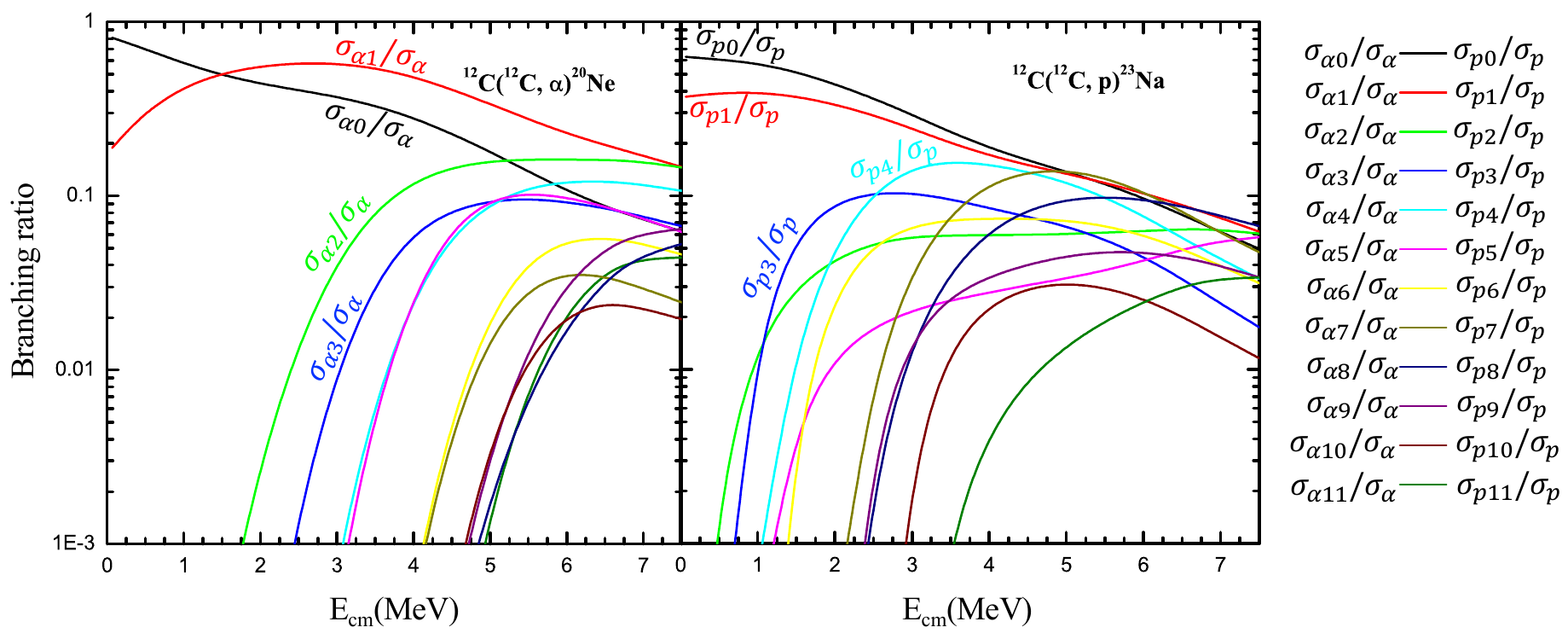}
\caption{The branching ratios, calculated using Talys \cite{koning2005}, for each state ($\alpha_{i}$, $p_{i}$, distinguished by color, where $i$=0,1,2,...,11) of the $\alpha$ and proton evaporation channels in the ${}^{12}$C+${}^{12}$C fusion reaction.}
\label{fig_branching_ratio_ap_all}
\end{figure*}

The decay of the carbon-fusion-made ${}^{24}$Mg compound nucleus with a given spin (0, 2, 4 or 6), is calculated using the statistical model code, Talys \cite{koning2005}. The numbers of experimentally known states considered in the present calculation are 50 for ${}^{23}$Na, 13 for ${}^{20}$Ne and 20 for ${}^{23}$Mg. The default level density and global optical model are used above the limit of the experimentally known states. After being weighted by the predicted spin population of the $^{24}$Mg compound nucleus shown in Fig.~\ref{spin_population_24Mg}, the branching ratios of each $\alpha$- and p- evaporation channel are derived and plotted in Fig.~\ref{fig_branching_ratio_ap_all}. The branching ratios for highly excited states of both channels fall rapidly with decreasing energy. In the $\alpha$-channel, the ground ($\alpha_0$) and first excited ($\alpha_1$) states dominate the total $\alpha$ evaporation cross sections for $\emph{E}_{\rm cm} \textless$3 MeV. However, in the p-channel, the situation is more complicated: besides the ground ($p_0$) and first excited ($p_1$) states, the $p_{2,3,4,5,6,7,8,9}$ also have appreciable contributions to the total proton evaporation cross sections.


\section{Comparisons between the calculations and the experimental results}
\label{sec_comparison}

\begin{figure*}[htbp]
\centering
\includegraphics[width=0.99\textwidth]{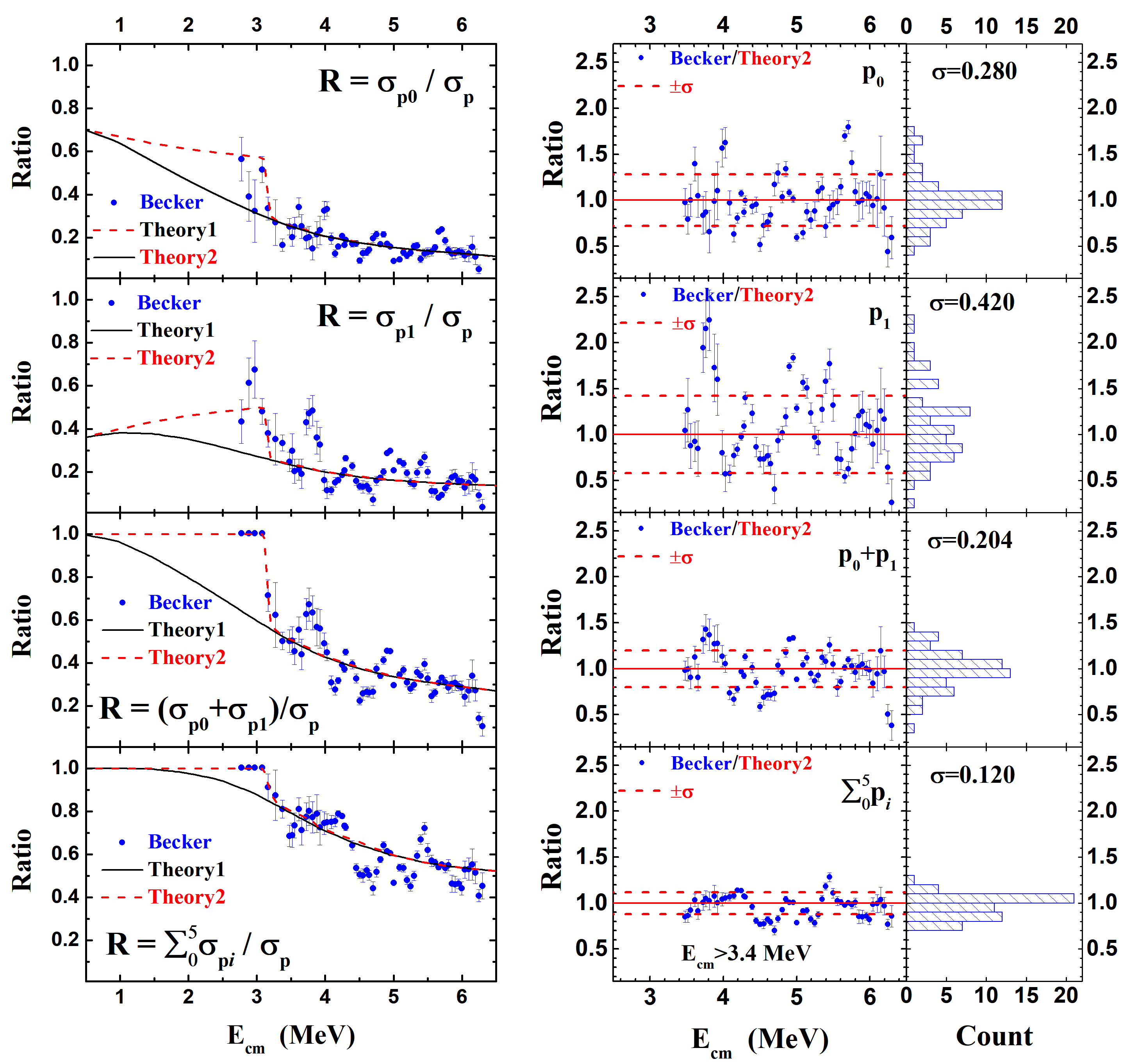}
\caption{A comparison of theoretical branching ratios with those of experimental data \cite{becker1981} for the p-channel. (left) The calculated $\sigma_{p_0}$/$\sigma_{p}$, $\sigma_{p_1}$/$\sigma_{p}$, ($\sigma_{p_0}+\sigma_{p_1}$)/$\sigma_{p}$, $\sum_{i=0}^5\sigma_{p_i}$/$\sigma_{p}$ ratios are displayed with experimental data from Becker $\emph{et al.}$ \cite{becker1981}. {\bf Theory1} is the present calculation by Talys while {\bf Theory2} takes into account the lack of experimental sensitivity to protons with energies below a certain cut-off value (see text for detail). (right) The values of Becker/Theory2 for the branching ratios shown on the left are calculated and displayed. The relevant statistics for each distribution are also displayed showing average value (Mean=1) and standard deviation ($\sigma$).}
\label{fig_ratio_p_channel}
\end{figure*}

\begin{figure*}[htbp]
\centering
\includegraphics[width=0.99\textwidth]{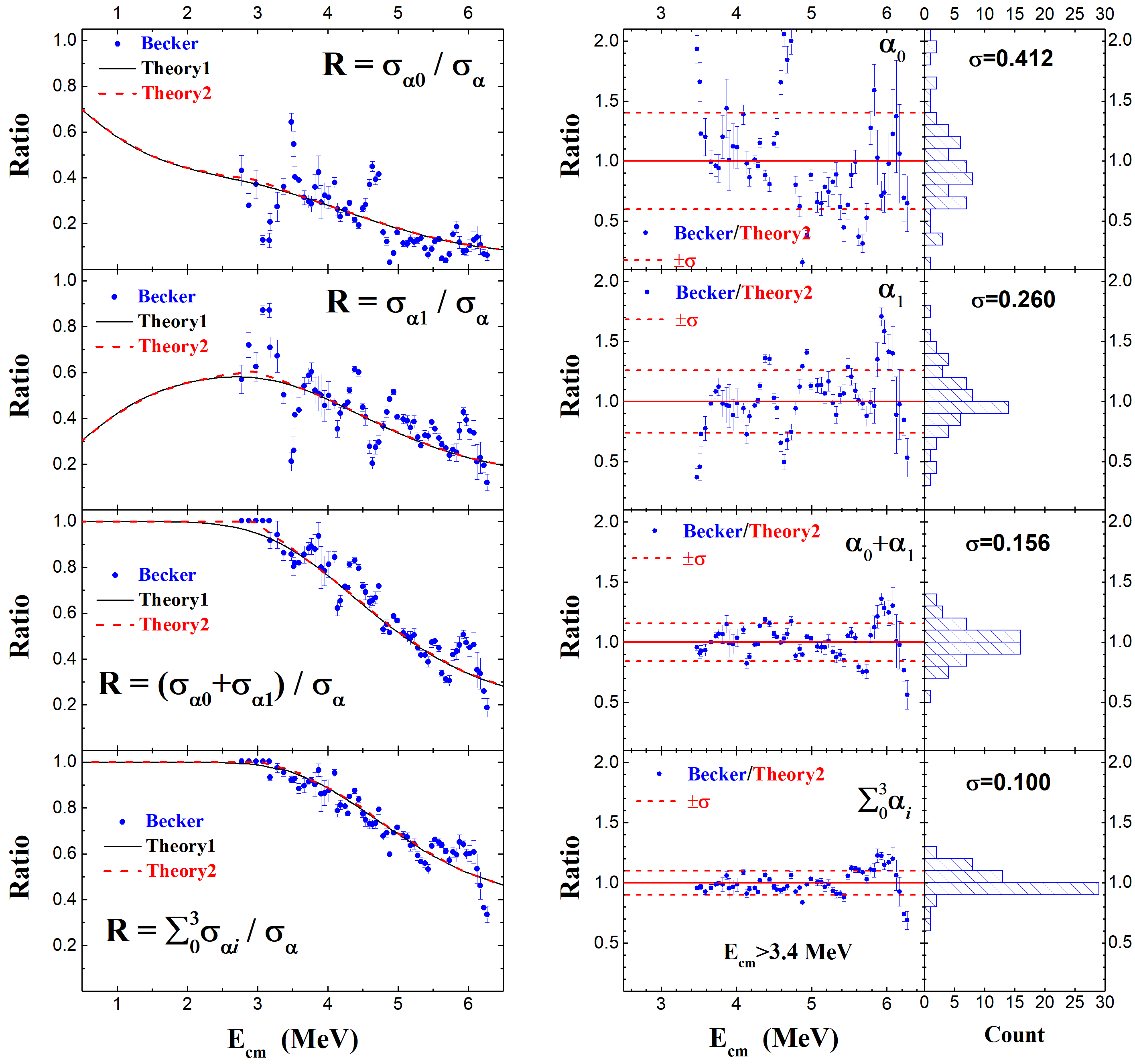}
\caption{A comparison of theoretical branching ratios with those of experimental data \cite{becker1981} for the $\alpha$-channel. (left ) The calculated $\sigma_{\alpha_0}$/$\sigma_{\alpha}$, $\sigma_{\alpha_1}$/$\sigma_{\alpha}$, ($\sigma_{\alpha_0}+\sigma_{\alpha_1}$)/$\sigma_{\alpha}$, and $\sum_{i=0}^3\sigma_{\alpha_i}$/$\sigma_{\alpha}$ ratios are displayed with experimental data from Becker $\emph{et al.}$ \cite{becker1981}. {\bf Theory1} is the present calculation by Talys while {\bf Theory2} takes into account the lack of experimental sensitivity to alphas with energies below a certain cut-off value (see text for detail). (right) The values of Becker/Theory2 for the branching ratios shown on the left are calculated and displayed. The relevant statistics for each distribution are also displayed showing average value (Mean=1) and standard deviation ($\sigma$).}
\label{fig_ratio_a_channel}
\end{figure*}

The calculated branching ratios are compared with the experimental values obtained by Becker $\emph{et al.}$ \cite{becker1981}, whom reported information of many separated states in both $\alpha$- and p- channels, including $\alpha_0$, $\alpha_1$, $\alpha_2$, $\alpha_3$, $\alpha_4$, $\alpha_5$, $\alpha_6$, $\alpha_7$, $\alpha_8$+$\alpha_9$, $\alpha_{10}$, $\alpha_{11}$, $\alpha_{12}$, and $p_0$, $p_1$, $p_2$, $p_3$, $p_4$+$p_5$, $p_6$, $p_7$, $p_8$+$p_9$, $p_{10}$, $p_{11}$, $p_{12}$, $p_{13}$, $p_{14}$+$p_{15}$+$p_{16}$. The $\sigma_{p_0}$/$\sigma_{p}$, $\sigma_{p_1}$/$\sigma_{p}$, ($\sigma_{p_0}+\sigma_{p_1}$)/$\sigma_{p}$, and $\sum_{i=0}^5\sigma_{p_i}$/$\sigma_{p}$ ratios of the present calculation are displayed with the experimental data of Becker $\emph{et al.}$ \cite{becker1981} on the left side of Fig.~\ref{fig_ratio_p_channel}. Similarly, the $\sigma_{\alpha_0}$/$\sigma_{\alpha}$, $\sigma_{\alpha_1}$/$\sigma_{\alpha}$, ($\sigma_{\alpha_0}+\sigma_{\alpha_1}$)/$\sigma_{\alpha}$, and $\sum_{i=0}^3\sigma_{\alpha_i}$/$\sigma_{\alpha}$ ratios are displayed with the corresponding experimental data on the left side of Fig.~\ref{fig_ratio_a_channel}. To compare the theory calculation of Talys with the actual particle measurement, we take the experimental cut-off energy for each observed state of Becker $\emph{et al.}$ \cite{becker1981} into account. The calculations of the branching ratios without considering cut-off energy are shown as solid lines and labeled {\bf Theory1}. In reality, the most highly-excited fusion residues are accompanied by very low energy protons and $\alpha$ particles that could not be seen by the experimental particle detectors and, thus, were missed in the measurements. These cut-off energies have been included in a second set of calculations shown as dashed lines and labeled {\bf Theory2}. 
The difference between Theory1 (ideal) and Theory2 (reality) arising from the missing channels is related to the detection threshold and/or background in the particle measurements. 
They are nearly the same for all energies in the relevant $\alpha$-channels, and for E$_{\rm cm}\textgreater$3.4 MeV in p-channels.

The present calculations (Theory 1 and Theory 2) describe a smoothly averaged trend for the experimental branching ratios obtained from Becker's data \cite{becker1981}.
The contribution of the ground state is about 40\% for $\sigma_{\alpha_0}$, and 30\% for $\sigma_{p_0}$ at $\emph{E}_{\rm cm}$=3 MeV. The branching ratios for $\sigma_{p_0}+\sigma_{p_1}$ and $\sigma_{\alpha_0}+\sigma_{\alpha_1}$ are larger than 50\% and 90\% at $\emph{E}_{\rm cm}$=3 MeV, respectively, and continuously increase with decreasing energy. There are significant fluctuations in the $p_0$, $p_1$, $\alpha_0$, and $\alpha_1$ channels due to strong resonances. 


To compare the trends of the branching ratios obtained from the particle spectroscopy measurement with the theoretical prediction, the values of Becker/Theory2 are calculated by dividing the experimental ratios with the theoretical predictions and displayed on the right side of Fig.~\ref{fig_ratio_p_channel} and Fig.~\ref{fig_ratio_a_channel}. Each theoretical ratio has been tuned using a renormalization factor, $f$, to achieve the best fit. For cases where the numerator in the ratio is also part of the denominator, such as $p_0$/$p_{tot}$, the scaled ratio is provided by numerator/(numerator+f(denominator-numerator)). Otherwise, $f$ is applied directly to the ratio. The values of f used ranged from 0.7 to 1.3. 

The statistical distribution of Becker/Theory2 for each branching ratio is analyzed in the energy range above 3.4 MeV in order to avoid the influence of missing channels. The distribution widths, which represent the fluctuation of the experimental values around the predictions, are summarized in Table \ref{tab_branching_ratio}. It is observed that the distribution widths of the Becker/Theory2 ratios are about 30\%(1$\sigma$) in the $p_0$, $p_1$, $\alpha_0$, and $\alpha_1$ channels. It is interesting to note that the fluctuations of data around theory become smaller as the branching ratios of the protons and $\alpha$ channels increase. 
In the p-channel, for example, the branching ratio of $\sum_{i=0}^5\sigma_{p_i}$/$\sigma_{p}$ is about 50\% to 70\% between 3.5 MeV$\textless\emph{E}_{\rm cm}\textless$6.5 MeV.
The fluctuation for $\sum_{i=0}^5\sigma_{p_i}$ is about $\pm$12\%(1$\sigma$) in contrast to the 30\% to 50\% (1$\sigma$) fluctuation observed in the $p_0$ and $p_1$ channels. A strong resonance is observed around $\emph{E}_{\rm cm}=$3.8 MeV in the summation of $p_0$ and $p_1$, but it disappears if we sum more proton channels up to $p_5$. As we increase the branching ratio of the observed proton channels, the fluctuation incurred by the resonance feature of the ${}^{12}$C+${}^{12}$C channel also becomes less.  
In the $\alpha$-channel, the branching ratio of $\sum_{i=0}^3\sigma_{\alpha_i}$/$\alpha$ is between 50\% and 100\% in the range 3.5 MeV$\textless\emph{E}_{\rm cm}\textless$6.5 MeV. The fluctuation for $\sum_{i=0}^3\sigma_{\alpha_i}$ is about $\pm$14\%(1$\sigma$) in contrast to the more than 30\%(1$\sigma)$ observed in $\alpha_0$ and $\alpha_1$. This fluctuation is expected to decrease considering the ratio ($\sigma_{\alpha_0}+\sigma_{\alpha_1}$)/$\alpha$ reaches nearly 100\% for ${E}_{\rm cm}\textless$3.5 MeV. 

It has been reported in ${}^{12}$C(${}^{13}$C,$p$)${}^{24}$Na that the average branching ratio is about 0.25 with a relative fluctuation of 14\%(1$\sigma$) \cite{zhang2020}. However, in order to reach a relative fluctuation down to 14\% in ${}^{12}$C+${}^{12}$C, the branching ratio has to be more than 0.5 for the proton channel and 0.7 for the $\alpha$ channel as shown in Table \ref{tab_branching_ratio}. This disparity may arise from significant differences in the compound nuclei, ${}^{24}$Mg and ${}^{25}$Mg \cite{jiang2018}. Compared to ${}^{12}$C+${}^{13}$C, the lower level density in the ${}^{12}$C+${}^{12}$C entrance and exit channels means the nuclear structure plays a stronger role which results in larger fluctuations in the branching ratio.


\begin{table}[]
\caption{\small The theoretical branching ratios at 4 MeV and their relative fluctuations in the range of 3.4 to 6 MeV.}
\label{tab_branching_ratio}
\begin{tabular}{|p{3.0cm} |p{1.6cm} |p{2.3cm}| }
\hline
Ratio                         & Value at 4 MeV & Relative fluctuation(1$\sigma$) \\ \hline
$\sigma_{p_0}$/$\sigma_p$               
                                  &  0.21               &  28\%                                 \\ \hline
$\sigma_{p_1}$/$\sigma_p$               
                                 &    0.21             &  42\%                       \\ \hline
($\sigma_{p_0}$+$\sigma_{p_1}$)/$\sigma_p$       
                                &  0.42                &  20\%                                \\ \hline
$\sum_{i=0}^5\sigma_{p_i}$/$\sigma_p$ 
                               &  0.71                 & 12\%                             \\ \hline 
$\sigma_{\gamma(440)}$/$\sigma_p$ 
                               &    0.48               &  14\%                             \\ \hline 
[$\sigma_{p_0}+\sigma_{\gamma(440)}$]/$\sigma_p$                            
                              &   0.69                &       8\%                       \\ \hline 
[$\sigma_{p_0}+\sigma_{\gamma(440)} \newline 
+\sigma_{\gamma(2391)}+\sigma_{\gamma(2640)} \newline
+\sigma_{\gamma(2982)}$]/$\sigma_p$ 
                              &   0.93                 &       3\%                       \\ \hline 
\hline
$\sigma_{\alpha_0}$/$\sigma_{\alpha}$               
                             &  0.24                  &  41\%                                 \\ \hline
$\sigma_{\alpha_1}$/$\sigma_{\alpha}$               
                             &    0.5                &  26\%                       \\ \hline
($\sigma_{\alpha_0}$+$\sigma_{\alpha_1}$)/$\sigma_{\alpha}$       
                             &  0.74                  &  16\%                                \\ \hline
$\sum_{i=0}^3\sigma_{\alpha_i}$/$\sigma_{\alpha}$ 
                            &  0.89                 & 10\%                             \\ \hline
$\sigma_{\gamma(1634)}$/$\sigma_{\alpha}$ 
                               &    0.67               &  16\%                             \\ \hline 
[$\sigma_{\alpha_0}+\sigma_{\gamma(1634)}$]/$\sigma_{\alpha}$                            
                              &   0.94                &       7\%                       \\ \hline 
\end{tabular}
\end{table}

\section{The connection between particle spectroscopy and $\gamma$-ray spectroscopy}
We use the statistical model result to estimate the branching ratios of the characteristic $\gamma$ rays, \emph{e.g.} 1634, 4967, 5621 keV for the $\alpha$-channel, and 440, 2076, 2391, 2640, 2982, 3848 keV for the p-channel. The cross sections for production of these characteristic $\gamma$ rays, which directly transit to the ground state ($\alpha_0$ or $p_0$), can be calculated by the expression

\begin{equation}
\sigma_\gamma=\sum_i f_i\sigma_i,
\label{eq1}
\end{equation}

\noindent where $\sigma_i$ are the partial cross sections for the $i^{th}$ excited state in the residual nucleus, $f_i$ represents the transit factors, of which the values are listed in Tables~\ref{tab1} and~\ref{tab2} for $\alpha$- and p- evaporation channels, respectively. They are deduced based on the $\gamma$ transition branching ratios from NNDC \cite{NNDC}.

\begin{table}[!htbp]
\caption{\small Numerical factors (\%) for the yields of the characteristic $\gamma$-rays from partial cross sections of the $\alpha$-evaporation channels.}
\label{tab1}
\begin{center}
\resizebox{0.28\textwidth}{!}{
\begin{tabular}{cccc}
\toprule[0.75pt] \noalign{\smallskip}
&$\sigma_{\gamma(1634)}$&$\sigma_{\gamma(4967)}$&$\sigma_{\gamma(5621)}$\\
\noalign{\smallskip}\hline\noalign{\smallskip}
$\sigma_{\alpha_1}$&100 & & \\
$\sigma_{\alpha_2}$&100 & & \\
$\sigma_{\alpha_3}$&99.4 &0.6 & \\
$\sigma_{\alpha_4}$&6.47 &0.002 &0.53 \\
$\sigma_{\alpha_8}$&76.6 &0.07 &0.13 \\
$\cdots$ & $\cdots$ & $\cdots$ & $\cdots$ \\
\noalign{\smallskip} \bottomrule[0.75pt]
\end{tabular}
}
\end{center}
\end{table}

\begin{table}[!htbp]
\caption{\small Numerical factors (\%) for the yields of the characteristic $\gamma$-rays from partial cross sections of the proton-evaporation channels.}
\label{tab2}
\begin{center}
\resizebox{0.5\textwidth}{!}{
\begin{tabular}{ccccccc}
\toprule[0.75pt] \noalign{\smallskip}
&$\sigma_{\gamma(440)}$&$\sigma_{\gamma(2076)}$&$\sigma_{\gamma(2391)}$&$\sigma_{\gamma(2640)}$&$\sigma_{\gamma(2982)}$&$\sigma_{\gamma(3848)}$\\
\noalign{\smallskip}\hline\noalign{\smallskip}
$\sigma_{p_1}$&100&&&&&\\
$\sigma_{p_2}$&91.8&8.2&&&&\\
$\sigma_{p_3}$&34.3&&65.7&&&\\
$\sigma_{p_4}$&&&&100&&\\
$\sigma_{p_5}$&97.1&2.9&&&&\\
$\sigma_{p_6}$&41.2&&0.2&&58.6&\\
$\sigma_{p_7}$&79.4&&0.9&19.5&0.3&\\
$\sigma_{p_8}$&66.4&5.0&0.004&4.5&1.2&22.9\\
$\sigma_{p_9}$&17.7&0.7&0.7&&1.35&\\
$\sigma_{p_{10}}$&2.8&&0.7&&&\\
$\sigma_{p_{11}}$&97.4&2.6&&&&\\
$\sigma_{p_{12}}$&80.9&1.8&0.01&&4.2&\\
$\sigma_{p_{13}}$&96.0&4.0&&&&\\
$\sigma_{p_{14}}$&29.1&&&&&\\
$\sigma_{p_{15}}$&43.1&0.08&&4.5&0.02&0.3\\
$\sigma_{p_{16}}$&67.1&&32.9&&&\\
$\vdots$&$\vdots$&$\vdots$&$\vdots$&$\vdots$&$\vdots$&$\vdots$\\
\noalign{\smallskip} \bottomrule[0.75pt]
\end{tabular}
}
\end{center}
\end{table}

The theoretical branching ratios of $\sigma_{\gamma(1634)}$/$\sigma_\alpha$ and $\sigma_{\gamma(440)}$/$\sigma_p$ are shown in Table \ref{tab3}. It should be noted that the 1634 keV transition of ${}^{20}$Ne is mixed with the 1636 transition of ${}^{23}$Na. The $\gamma$-spectroscopy can not resolve these two $\gamma$-rays due to the Doppler broadening. This contribution can be estimated based on the observed yield of the 440 keV transition from ${}^{23}$Na and the predicted ratio of $\sigma_{\gamma(1636)}$/$\sigma_{\gamma(440)}$. 

\begin{table}[!htbp]
\caption{\small Theoretical branching ratios predicted using the statistical model.}
\label{tab3}
\begin{center}
\resizebox{0.5\textwidth}{!}{
\begin{tabular}{ccccccccc}
\toprule[0.75pt] \noalign{\smallskip}
\multicolumn{1}{c}{$\emph{E}_{\rm cm}$} & \multicolumn{1}{|c|}{$\sigma_{p_0}$/$\sigma_p$} & $\sigma_{\gamma(440)}$/$\sigma_p$ & \multicolumn{1}{|c|}{$\sigma_{\alpha_0}$/$\sigma_\alpha$} & $\sigma_{\gamma(1634)}$/$\sigma_\alpha$ & \multicolumn{1}{|c|}{$\sigma_{\gamma(1636)}$/$\sigma_{\gamma(440)}$}  \\
\noalign{\smallskip}\hline\noalign{\smallskip}
0.5 & 0.6986 & 0.3432 & 0.6798 & 0.3260 & 0.0063 \\
0.6 & 0.6868 & 0.3518 & 0.6570 & 0.3490 & 0.0112 \\
0.8 & 0.6638 & 0.3683 & 0.6104 & 0.3960 & 0.0201 \\
1   & 0.6378 & 0.3847 & 0.5665 & 0.4400 & 0.0288 \\
1.2 & 0.6025 & 0.3988 & 0.5336 & 0.4731 & 0.0392 \\
1.4 & 0.5673 & 0.4125 & 0.5016 & 0.5051 & 0.0489 \\
1.5 & 0.5494 & 0.4181 & 0.4879 & 0.5187 & 0.0536 \\
1.6 & 0.5322 & 0.4214 & 0.4777 & 0.5290 & 0.0587 \\
1.8 & 0.4977 & 0.4277 & 0.4572 & 0.5494 & 0.0687 \\
2   & 0.4640 & 0.4323 & 0.4383 & 0.5683 & 0.0786 \\
2.2 & 0.4321 & 0.4336 & 0.4234 & 0.5831 & 0.0890 \\
2.4 & 0.4005 & 0.4344 & 0.4091 & 0.5973 & 0.0996 \\
2.5 & 0.3851 & 0.4357 & 0.4022 & 0.6040 & 0.1051 \\
2.6 & 0.3700 & 0.4374 & 0.3953 & 0.6108 & 0.1108 \\
2.8 & 0.3408 & 0.4400 & 0.3814 & 0.6242 & 0.1224 \\
3   & 0.3120 & 0.4452 & 0.3672 & 0.6370 & 0.1338 \\
3.2 & 0.2867 & 0.4523 & 0.3522 & 0.6482 & 0.1455 \\
3.4 & 0.2613 & 0.4593 & 0.3369 & 0.6597 & 0.1567 \\
3.5 & 0.2508 & 0.4630 & 0.3289 & 0.6626 & 0.1624 \\
3.6 & 0.2412 & 0.4670 & 0.3198 & 0.6642 & 0.1676 \\
3.8 & 0.2219 & 0.4750 & 0.3015 & 0.6678 & 0.1780 \\
4   & 0.2048 & 0.4826 & 0.2832 & 0.6664 & 0.1877 \\
4.2 & 0.1922 & 0.4870 & 0.2621 & 0.6570 & 0.1968 \\
4.4 & 0.1793 & 0.4927 & 0.2420 & 0.6465 & 0.2049 \\
4.5 & 0.1743 & 0.4937 & 0.2310 & 0.6397 & 0.2089 \\
4.6 & 0.1696 & 0.4940 & 0.2208 & 0.6308 & 0.2127 \\
4.8 & 0.1608 & 0.4958 & 0.2003 & 0.6136 & 0.2194 \\
5   & 0.1527 & 0.4962 & 0.1810 & 0.5952 & 0.2267 \\
5.2 & 0.1465 & 0.4955 & 0.1637 & 0.5756 & 0.2339 \\
5.4 & 0.1400 & 0.4954 & 0.1454 & 0.5568 & 0.2410 \\
5.5 & 0.1374 & 0.4952 & 0.1387 & 0.5477 & 0.2454 \\
5.6 & 0.1351 & 0.4948 & 0.1325 & 0.5389 & 0.2505 \\
5.8 & 0.1299 & 0.4946 & 0.1196 & 0.5219 & 0.2606 \\
6   & 0.1250 & 0.4950 & 0.1086 & 0.5066 & 0.2714 \\
6.2 & 0.1202 & 0.4966 & 0.1003 & 0.4952 & 0.2830 \\
6.4 & 0.1149 & 0.4982 & 0.0920 & 0.4842 & 0.2957 \\
6.5 & 0.1120 & 0.4991 & 0.0878 & 0.4785 & 0.3024 \\
\noalign{\smallskip} \bottomrule[0.75pt]
\end{tabular}
}
\end{center}
\end{table}


\section{The S$^*$ factors of ${}^{12}$C(${}^{12}$C,$p$)${}^{23}$N\MakeLowercase{a} and ${}^{12}$C(${}^{12}$C,$\alpha$)${}^{20}$N\MakeLowercase{e} at $\emph{E}_{\rm cm}>$2.7 M\MakeLowercase{e}V}

The statistical model calculation provides branching ratios to convert the observed particle or characteristic $\gamma$-ray S$^*$ factors into the S$^*$ factors of ${}^{12}$C(${}^{12}$C,$p$)${}^{23}$Na and ${}^{12}$C(${}^{12}$C,$p$)${}^{23}$Na. Two principles are followed to reduce the systematic error of the statistical model. First, the total S$^*$ factor of the proton (alpha) channel is only calculated from the corresponding S$^*$ factors observed by particle spectroscopy or characteristic-$\gamma$ rays to minimize the influences incurred by the mismatched optical potentials of the $p$ and $\alpha$ channels. Second, as many observed channels as possible are included which increases the branching ratio and results in less systematic fluctuations.

\begin{figure*}[htbp]
\centering
\includegraphics[width=0.99\textwidth]{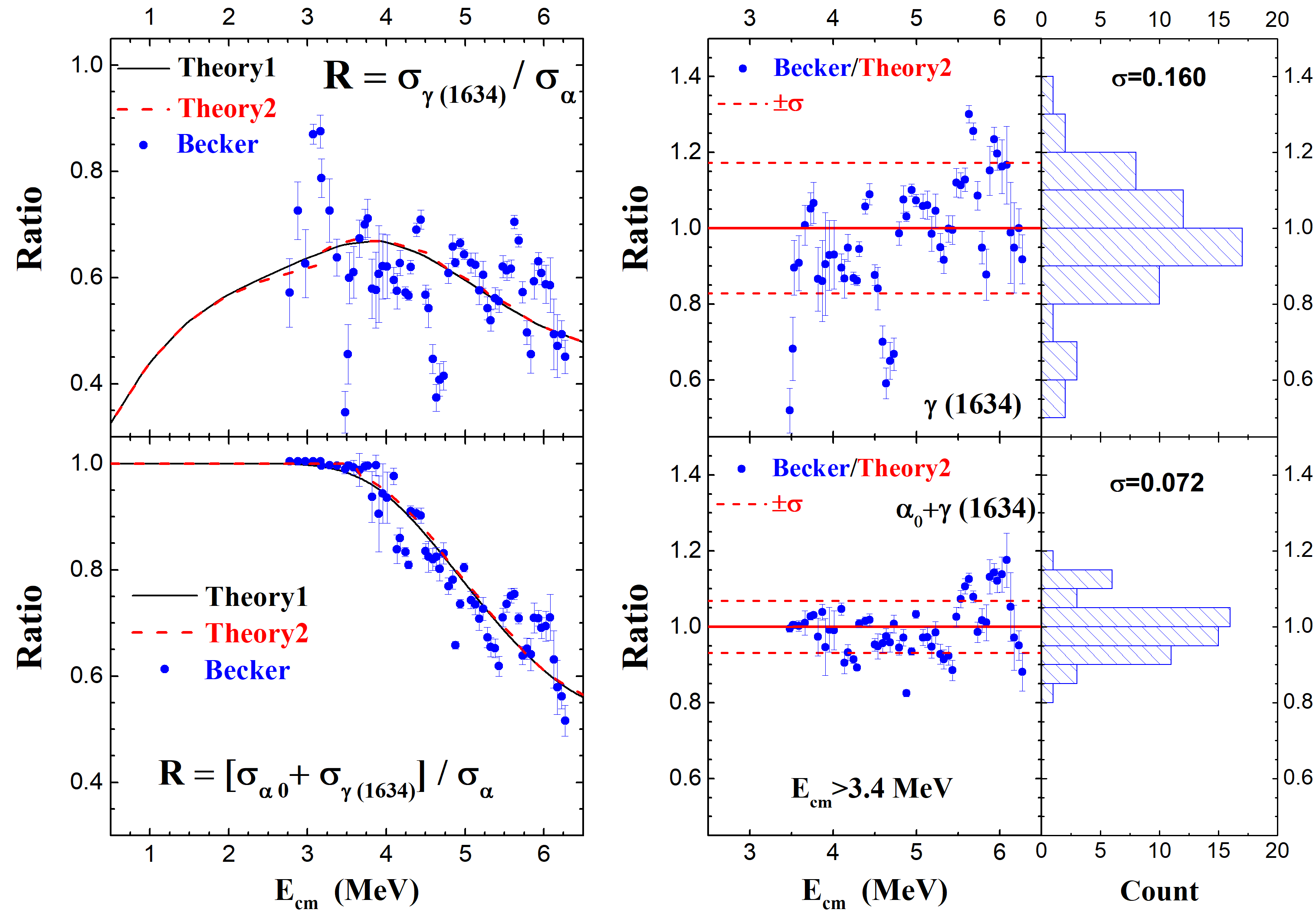}
\caption{A comparison of the branching ratios calculated with experimental data measured by Becker $\emph{et al.}$ \cite{becker1981} for $\alpha_0$ and the 1634 keV characteristic $\gamma$ ray. (left) The $\sigma_{\gamma(1634)}$/$\sigma_{\alpha}$ and [$\sigma_{\alpha_0}$ + $\sigma_{\gamma(1634)}$]/$\sigma_{\alpha}$ ratios are shown. {\bf Theory1} is the present calculation by Talys, {\bf Theory2} takes into account the experimental cut-off energies of Becker $\emph{et al.}$ \cite{becker1981}, as discussed in the text. (right) The values of Becker/Theory2 for the ratios shown on the left are calculated and displayed. The relevant statistics are also shown with average value (Mean=1) and standard deviation ($\sigma$).}
\label{fig_ratio_a0_1634}
\end{figure*}

\begin{figure*}[!htbp]
\centering
\includegraphics[width=0.99\textwidth]{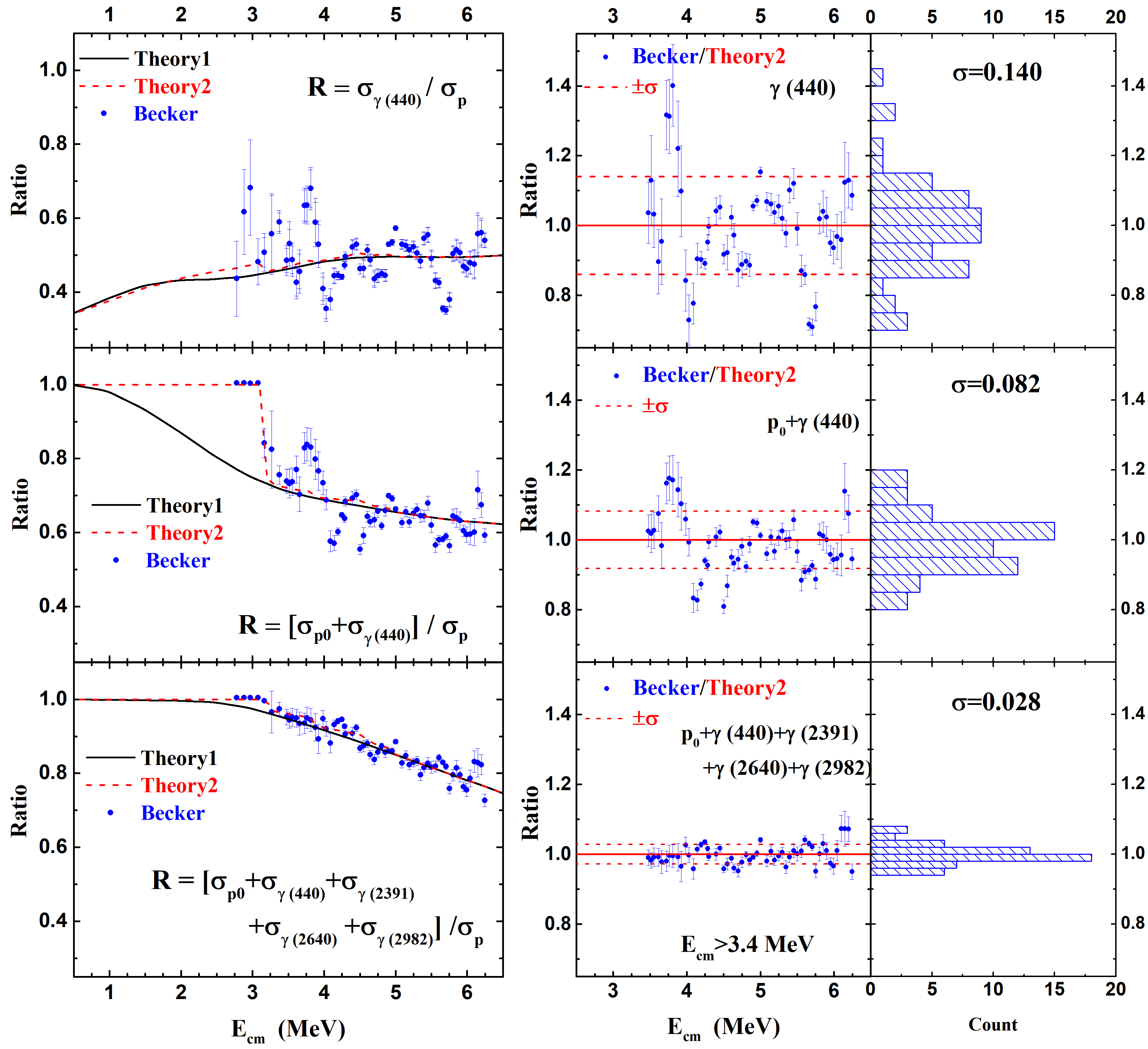}
\caption{A comparison of the branching ratios calculated with experimental data measured by Becker $\emph{et al.}$ \cite{becker1981} for $p_0$, 440 keV characteristic and other $\gamma$ rays. (left) The $\sigma_{\gamma(440)}$ / $\sigma_{p}$, [$\sigma_{p_0}$ + $\sigma_{\gamma(440)}$] / $\sigma_{p}$, and [$\sigma_{p_0}$ + $\sigma_{\gamma(440)}$ + $\sigma_{\gamma(2391)}$ + $\sigma_{\gamma(2640)}$ + $\sigma_{\gamma(2982)}$] / $\sigma_{p}$ ratios are shown. {\bf Theory1} is the present calculation by Talys, {\bf Theory2} takes into account the experimental cut-off energies of Becker $\emph{et al.}$ \cite{becker1981}, as discussed in the text. (right) The values of Becker/Theory2 for the ratios shown on the left are calculated and displayed. The relevant statistics are also shown with average value (Mean=1) and standard deviation ($\sigma$).}
\label{fig_ratio_p0_440}
\end{figure*}

For the measurements using the characteristic $\gamma$-ray method, we add the ground state transitions (${}^{12}$C(${}^{12}$C,$p_0$)${}^{23}$Na and ${}^{12}$C(${}^{12}$C,$\alpha_0$)${}^{20}$Ne obtained with particle spectroscopy to the observed $\gamma$-ray S$^*$ factors. The theoretical and experimental branching ratios are shown in Fig. \ref{fig_ratio_a0_1634} and Fig. \ref{fig_ratio_p0_440}. The inclusion of the ground state channels ($p_0$, $\alpha_0$) is important especially for $\emph{E}_{\rm cm}<$3.5 MeV because these ground state channels are expected to contribute significantly as shown in Fig. \ref{fig_ratio_p_channel} and Fig. \ref{fig_ratio_a_channel}. For the $\alpha$ channel, the sum of $\sigma_{\alpha 0}$ and $\sigma_{\gamma(1634)}$ contributes about 60\% $\sim$ 90\% of the ${}^{12}$C(${}^{12}$C,$\alpha$)${}^{20}$Ne cross sections at 3.4 $\sim$ 6.0 MeV in the center of mass frame. The fluctuation of the experimental [$\sigma_{\alpha_0}$ + $\sigma_{\gamma(1634)}$]/$\sigma_{\alpha}$ around the predicted ratio reflects the systematic error of the latter which is estimated to be 7.2\% based on the experimental data in the energy range from 3.4 to 6.0 MeV. The fluctuation is expected to be smaller and eventually vanish at stellar energies as the summation becomes equal to the total cross sections of the  ${}^{12}$C(${}^{12}$C,$\alpha$)${}^{20}$Ne at these energies. Therefore, only measuring these two components is sufficient for determining the ${}^{12}$C(${}^{12}$C,$\alpha$)${}^{20}$Ne S$^*$ factor in the important energy range. The proton channel is a little more complicated. The sum of $\sigma_{p_0}$ and $\sigma_{\gamma(440)}$ contributes nearly 70\% of the ${}^{12}$C(${}^{12}$C,$p$)${}^{23}$Na S$^*$ factor at $\emph{E}_{\rm cm}$= 4 MeV. The systematic error of the theoretical [$\sigma_{p_0}$ + $\sigma_{\gamma(440)}$] / $\sigma_{p}$ ratio is estimated to be 8.2\%. As the ratio of [$\sigma_{p_0}$ + $\sigma_{\gamma(440)}$] / $\sigma_{p}$ increases from 70\% $\sim$ 90\% at astrophysical energies, the systematic uncertainty is expected to get smaller. By including more transitions such as 2391 keV, 2640 keV and 2982 keV of ${}^{23}$Na, the ratio of the sum of these observable channels to the ${}^{12}$C(${}^{12}$C,$p$)${}^{23}$Na becomes more than 90\% at all energies shown in Fig. \ref{fig_ratio_p0_440}, and the systematic error is less than 3\% which is mainly limited by the experimental uncertainty of the branching ratio.

\begin{figure*}[!htbp]
\centering
\includegraphics[width=1.0\textwidth]{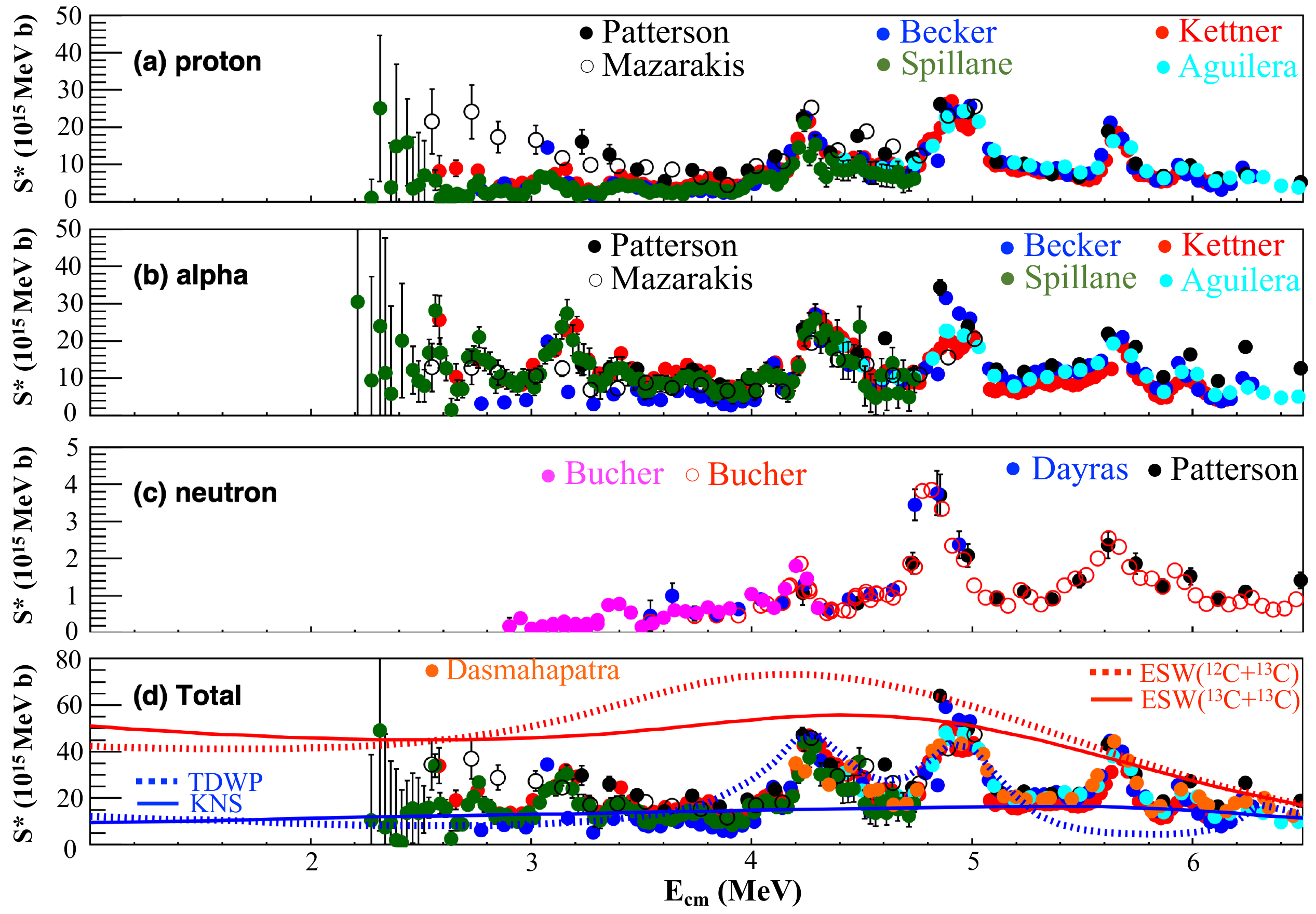}
\caption{S$^*$ factor of the proton, alpha, neutron channels and the sum of the three channels after correcting for missing channels.}
\label{fig_sfactor_p_a}
\end{figure*}

The total p- and $\alpha$- S$^*$(E) factors from charged-particle measurements of Patterson $\emph{et al.}$ \cite{patterson1969}, Mazarakis and Stephens \cite{mazarakis1973}, and Becker $\emph{et al.}$ \cite{becker1981} are shown in Fig. \ref{fig_sfactor_p_a}. The Becker $\emph{et al.}$ data were corrected to account for the missing channels by applying the ratios in Fig.~\ref{fig_ratio_p_channel} and Fig.~\ref{fig_ratio_a_channel}. The data of Mazarakis $\emph{et al.}$ \cite{mazarakis1973} and Kettner $\emph{et al.}$ \cite{kettner1977} are shifted by 100 keV and 30 keV respectively to match the resonances measured by Aguilera $\emph{et al.}$ \cite{aguilera2006} and Spillane $\emph{et al.}$ \cite{spillane2007}. 
 

Based on the ground state data from Becker $\emph{et al.}$ \cite{becker1981} and the calculated ratios for [$\sigma_{\alpha 0}$ + $\sigma_{\gamma(1634)}$]/$\sigma_{\alpha}$ and [$\sigma_{p0}$ + $\sigma_{\gamma(440)}$] / $\sigma_{p}$ (Fig.\ref{fig_ratio_a0_1634} and \ref{fig_ratio_p0_440}), the data of Kettner $\emph{et al.}$ \cite{kettner1977}, Aguilera $\emph{et al.}$ \cite{aguilera2006} and Spillane $\emph{et al.}$ \cite{spillane2007} have been corrected. 

\begin{figure*}[!htbp]
\centering
\includegraphics[width=1.0\textwidth]{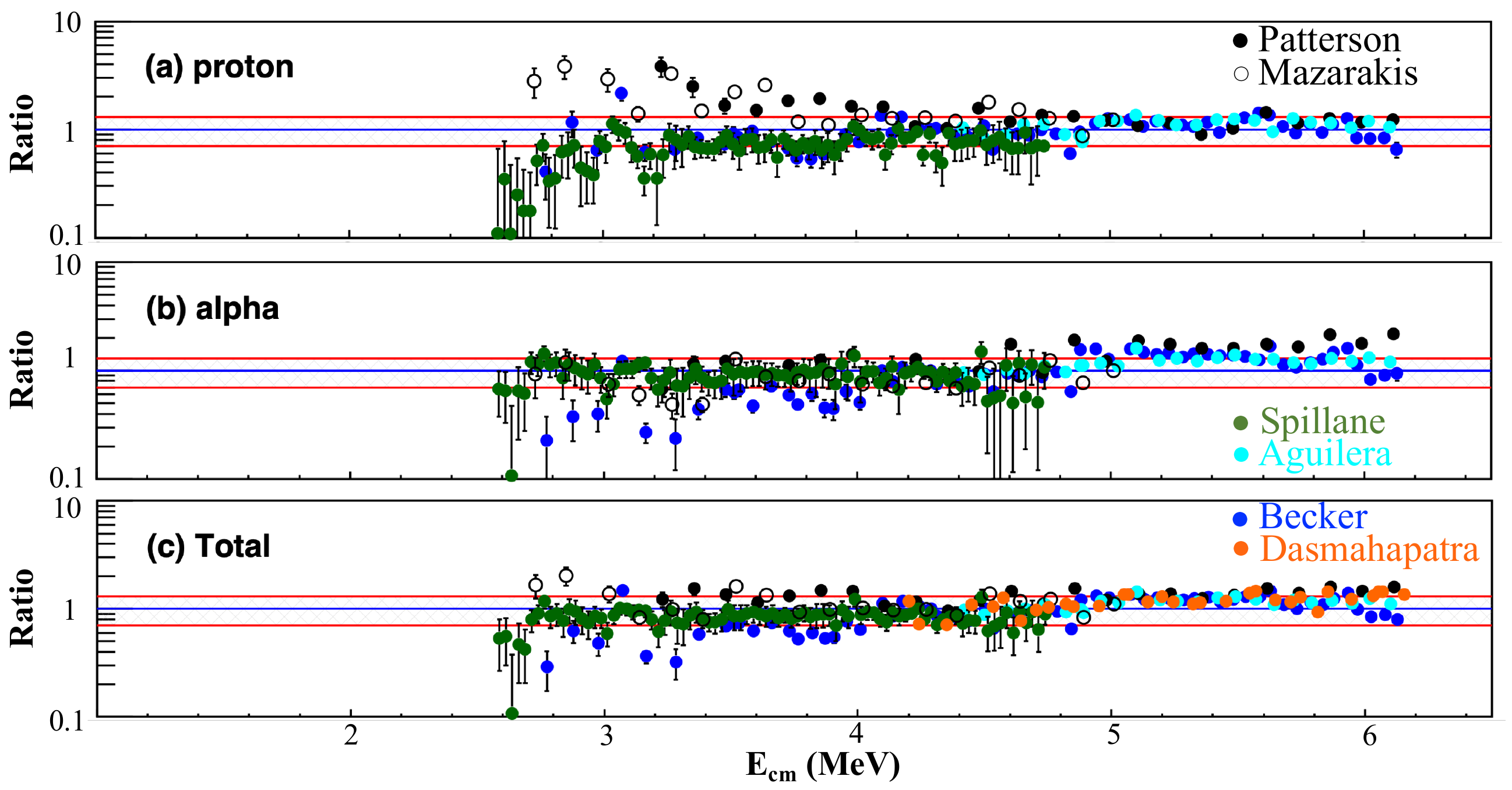}
\caption{Ratios of the various S$^*$ factors to the baseline S$^*$ factors, which were measured by Kettner $\emph{et al.}$ after correcting for the missing channels. The means and errors of the baseline data sets are interpolated in the calculations of ratios. The shaded areas shown in the ratio plots correspond to $\pm$30\% deviations.}
\label{fig_ratio_new}
\end{figure*}

Using the data of Kettner $\emph{et al.}$  \cite{kettner1980} as the baseline, the ratios of the S$^*$ factors of the proton and alpha channels and the total S$^*$ factors are computed for each data set and shown in Fig. \ref{fig_ratio_new}.  

For the proton channel, the S$^*$ factors obtained with $\gamma$-spectroscopy (Spillane, Aguilera and Kettner) agree with the S$^*$ factors obtained with particle-spectroscopy (Becker, Patterson and Mazarakis) within $\pm$ 30\% at energies above 4 MeV. This observation shows the importance and effectiveness of the correction done in the current paper. Some systematic deviations are observed in the range of 5 to 6 MeV, possibly arising from systematic errors in some of the experiments which might be improved in future experiments. At energies below 4 MeV, the result of Spillane is about (20-30)\% lower than the baseline data (Kettner). Since both experiments were done with $\gamma$-ray spectroscopy and corrected by the same branching ratio, this difference can only arise from experimental systematic errors. The three particle spectroscopy experiments, Becker, Patterson and Mazarakis, disagree at energies below 4.2 MeV. The results obtained from the measurements of Patterson and Mazarakis are nearly a factor of 4 higher than the result obtained from the measurement of Becker. Yet the S$^*$ factor of Spillane agrees with the one from Becker even though these two measurements were done using different methods. It is well known that the deuterium impurity in carbon targets may contribute a background in the proton spectrum via the $d$(${}^{12}$C,$p$)${}^{13}$C reaction. Since the measurement of 440 keV is less affected by the $\gamma$-ray from the reaction incurred by the deuterium impurity, the S$^*$ factors obtained with $\gamma$-ray spectroscopy seem to be more reliable. The agreement between the results of Spillane and Becker suggests that the impurity contribution might be better controlled in the measurement of Becker.  Future measurements are desired at energies below 4 MeV to resolve the differences among the various ${}^{12}$C(${}^{12}$C,$p$)${}^{23}$Na data sets. 

For the alpha channel, the situation is better. All the data sets seem to agree with each other within $\pm$30\% with two exceptions. The result of Patterson is higher than others at energies above 5.6 MeV. This deviation can be explained by considering the ${}^{16}$O+2$\alpha$(or ${}^8$Be) channel. Experiment shows that the branching ratio of the 2$\alpha$ channel increases as the $\emph{E}_{\rm cm}$ increases \cite{cujec1989}. Possibly limited by the Q-value resolution, Patterson might mix the 2$\alpha$ channel with the 1$\alpha$ channel, leading to a larger ${}^{12}$C(${}^{12}$C,$\alpha$)${}^{20}$Ne S$^*$ factor. The other exception is that the result obtained with the Becker measurement is significantly lower than all the other results at energies less than 3.4 MeV. It has been shown in the previous sections that the theoretical branching ratios agree well with the experimental branching ratios based on the Becker measurement. The large deviation observed at energies less than 3.4 MeV seems to suggest that there are missing $\alpha$ channels in the Becker measurement. Yet somehow the other measurements by Mazarakis and Patterson were able to catch all the major channels. 

For the total S$^*$ factors, a reasonable agreement is observed at energies above 4 MeV. The total S$^*$ factors obtained with the measurements of Aguilera, Becker and Patterson mostly agree with each other within 20\% in the three channels ranging between 4.4 to 6.1 MeV. The measurement using total absorption $\gamma$-spectroscopy agrees with the other measurements as well \cite{dasmahapatra1982}. However all three data sets are (20-30)\% higher than the total S$^*$ factors of Kettner in the range of 4.95 to 5.6 MeV. This systematic deviation may come from experimental systematic errors. At energies below 4.7 MeV, the five data sets, Kettner, Spillane, Becker, Mazarakis and Patterson agree with each other within $\pm$30\% other than a few points obtained from two of the particle spectroscopy measurements, Becker and Mazarakis below 3.6 MeV. The large deviation observed among the various S$^*$ factors of the proton channel is diluted, indicating that the alpha channel gets stronger at lower energies. 



\section{Comparison of the indirect and direct measurements at $\emph{E}_{\rm cm}<$2.7 M\MakeLowercase{e}V}

\begin{figure}[!htbp]
\centering
\includegraphics[width=0.5\textwidth]{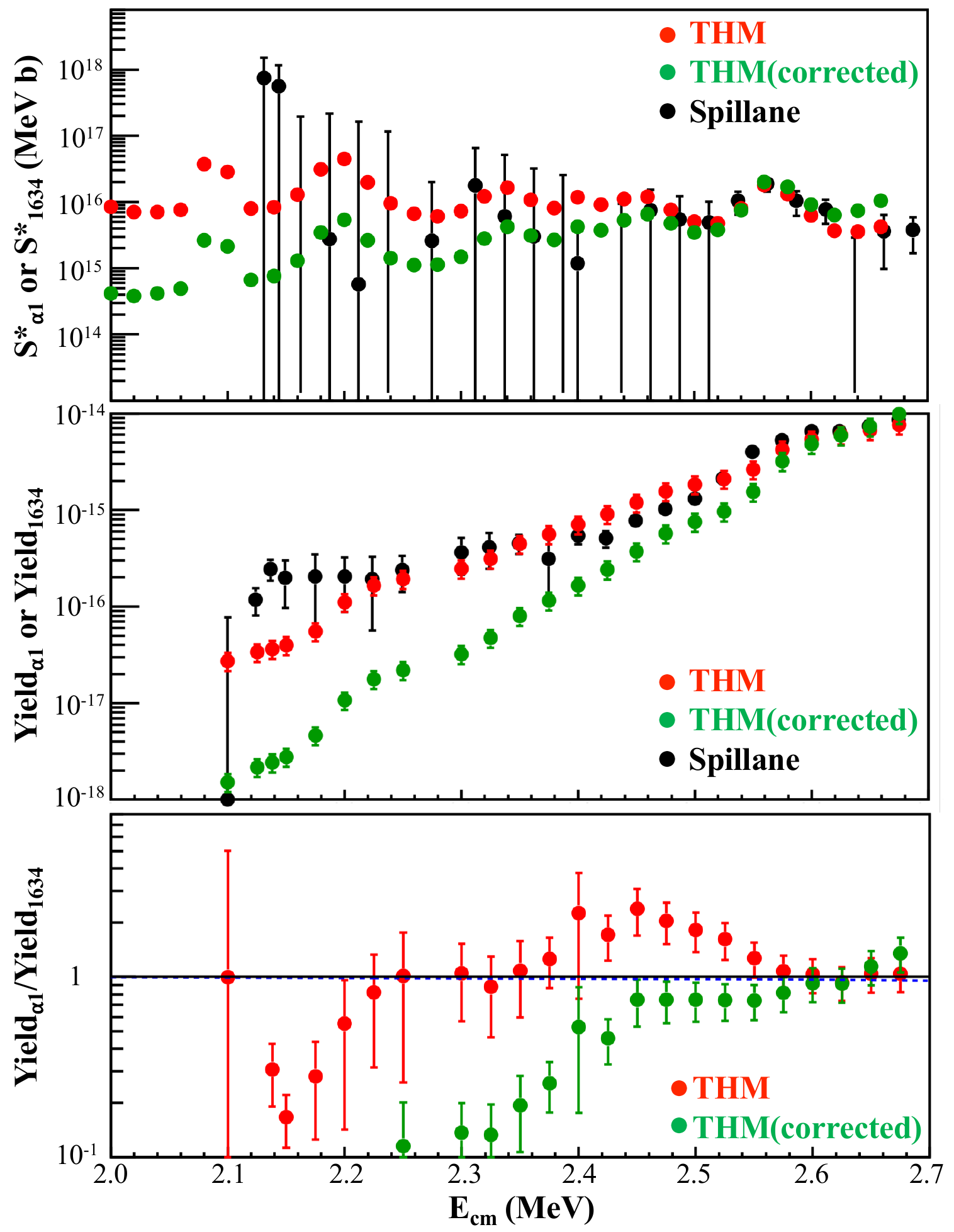}
\caption{(Top) S$^*$ factors of the $\alpha_1$ channel obtained with THM(red), corrected THM(green) and the 1634 keV transition measured by Spillane $\emph{et al.}$ (black);  (Middle) The thick target yield measured by Spillane $\emph{et al.}$ (black) and the calculated thick target yields based on the THM and the corrected THM data sets; (Bottom) The thick target ratios of the calculated yields based on THM (red) or corrected THM (green) to the one measured by Spillane $\emph{et al.}$.}
\label{fig_a1_smod_yield_ratio}
\end{figure}

The Trojan Horse Method (THM) has been done by Tumino $\emph{et al.}$ \cite{tumino2018} in the range of 0.8 MeV to 2.7 MeV to provide an extrapolation for the S$^*$ factor of ${}^{12}$C+${}^{12}$C. This method provides the shape of the energy dependence of the S$^*$ factor. The absolute value is fixed by normalizing the THM result to the direct measurement. It has been pointed out that the significant rise in S$^*$-factor observed in Ref. \cite{tumino2018} appeared mainly due to the invalid plane-wave approach used in the region where the Coulomb interaction is crucial. After applying a more general theory developed in Ref. \cite{akram2011}, the S$^*$-factor found in Ref. \cite{tumino2018} is greatly reduced \cite{akram2019}. The two versions of S$^*$$_{\alpha_1}$ factors are shown in Fig. \ref{fig_a1_smod_yield_ratio}. A comparison of the direct measurements of ${}^{12}$C+${}^{12}$C fusion and the indirect THM data was presented \cite{beck2020}.

It has been reported in the measurement of Becker that the ratio of $\sigma_{\alpha2}$/$\sigma_{\alpha1}$ ratio is less than 2\% at 3.18 MeV and lower energies. Our statistical model predicts this ratio to be below 6\% at energies less than 3 MeV. Therefore, the $\alpha_2$ contribution can be ignored and the full 1634 keV $\gamma$ ray yield considered equal to the $\alpha_1$ channel. The THM measurement normalized the $\alpha_1$ channel to the total S$^*$ factor of the $\alpha$ channel which includes the $\alpha_0$ channel.  Therefore we have to renormalize the S$^*$$_{\alpha_1}$ factor obtained with THM to the directly measured S$^*$$_{1634}$ factor by Spillane $\emph{et al.}$ using the resonance at $\emph{E}_{\rm cm}$=2.567 MeV.  

A comparison of the directly measured S$^*$ factor to the two versions of the S$^*$ factors obtained with the THM are shown in Fig. \ref{fig_a1_smod_yield_ratio}. Although some discrepancies have been observed at energies below 2.7 MeV, the large error bars of the Spillane measurement prevent a clear conclusion. The origin of the large errors comes in part from low statistics and from limitations caused by the background level. Another contribution comes from the differentiation process to get the cross section from the measured thick target yield. To avoid this uncertainty, we choose to convert the two versions of the THM S$^*$ factors into the thick target yields using the formula in Ref. \cite{notani2012} and compare them with the measured thick target yield of the 1634 keV transition done by Spillane $\emph{et al.}$. The comparison of the thick target yields is shown in Fig. \ref{fig_a1_smod_yield_ratio}.

The ratios of the THM thick target yields to the directly measured thick target yield are also shown at the bottom of Fig. \ref{fig_a1_smod_yield_ratio}. There are two clear disagreements between the S$^*$ factors obtained with indirect and direct measurements. The first disagreement is around 2.14 MeV where a strong resonance was claimed. Up to now, there has been no precise direct measurement able to confirm the existence of this resonance. Moreover it is clear that the two THM thick target yields are lower than Spillane's thick target yield and the strong resonance claimed at 2.14 MeV does not appear in the indirect measurement. Another clear disagreement happens between 2.4 to 2.6 MeV. The THM result is nearly a factor of 2 higher than the direct measurement having a reduced $\chi^2$ of 2.88. The calculated thick target yield using the THM S$^*$ factor corrected with the distorted wave approximation is about 20\% lower than the direct measurement at energies between 2.45 MeV to 2.6 MeV, but this yield becomes much less than the direct measurement at lower energies. Based on this comparison of thick target yields, we may conclude that neither THM result agrees with the direct measurement at energies below 2.7 MeV. More and better direct measurements with higher precision are needed at energies below 2.7 MeV to guide the development of the indirect measurement method.

\section{Extrapolation of the total ${}^{12}$C+${}^{12}$C S$^*$ factor}
\label{sec_extrapolation_S$^*$}

The astrophysically interesting energy range covers from a few tens keV up to 3 MeV. Extrapolating the
averaged ${}^{12}$C+${}^{12}$C fusion cross section down to stellar energies is inevitably needed. The complicated resonance structure in ${}^{12}$C+${}^{12}$C and the lack of reliable measurements at lower energies prevent us from drawing a clear conclusion \cite{fowler1984,denisov2019,denisov2010}. The standard reaction rate (CF88 \cite{caughlan1988}) was established by using a constant S$^*$(E) based on the square well penetration factor. The trend of the predicted S$^*$ factor agrees with later theoretical calculations, such as the coupled channel calculation (CC-M3Y+Rep), barrier penetration model based on the S$\tilde{a}$o Paulo potential (SPP), the Krappe-Nix-Sierk potential (KNS) or Equivalent Square Well potential (ESW), density-constrained time dependent Hartree-Fock method (DC-TDHF), and wave-packet dynamics (TDWP).

The hindrance model predicts that the ${}^{12}$C+${}^{12}$C S-factor reaches its maximum around $\emph{E}_{\rm cm}$=3.68 MeV \cite{esbensen2011}. At lower energies, this model predicts a rapid drop in the S-factor leading to a reduced reaction rate that is many orders of magnitude smaller than the standard rate used for astrophysical modeling (see Fig. \ref{fig_rate_ratio}). A satisfactory description of both the $^{12}$C+$^{12}$C and $^{12}$C+$^{13}$C cross sections with one set of assumptions and parameters is mandatory for any global model being tested. The recent precise measurement of the ${}^{12}$C+${}^{13}$C at deep sub-barrier energies clearly rules out the existence of the astrophysical S-factor maximum predicted by the phenomenological hindrance model, while confirming the trend of the S$^*$ factor towards lower energies predicted by other models, such as CC-M3Y+Rep, DC-TDHF, KNS, SPP and ESW \cite{zhang2020}. A recent TDHF calculation also claims the absence of hindrance in $^{12}$C+$^{12}$C \cite{godbey2019}. 

The strong correlation among the carbon isotope systems provides a great opportunity to establish an upper limit for the ${}^{12}$C+${}^{12}$C S$^*$ factor by using models constrained by ${}^{12}$C+${}^{13}$C or ${}^{13}$C+${}^{13}$C \cite{notani2012}. Later, this correlation was explained by the ANL group using the large differences of level densities between ${}^{12}$C+${}^{12}$C and other carbon isotope systems \cite{jiang2013}. The upper limit obtained with ${}^{12}$C+${}^{13}$C has been reported in Ref. \cite{zhang2020}. In the present paper, we fit the ${}^{13}$C+${}^{13}$C data using the ESW model. By following the suggestion of Esbensen $\emph{et al.}$ \cite{esbensen2011}, we scale this original data by a factor of 1.2 in our fitting. The best fit is achieved with the parameters V=-4.01093 MeV, W=1.02877 MeV, R=7.35012 fm. The upper limit of ${}^{12}$C+${}^{12}$C is obtained by scaling R with (12/13)$^{(1/3)}$. The two upper limits are shown in Fig. \ref{fig_sfactor_p_a} (d). Although both limits provide a good upper bound for the total S$^*$ factor of ${}^{12}$C+${}^{12}$C, there are minor differences between them. The upper limit obtained with ${}^{12}$C+${}^{13}$C is higher than the upper limit obtained with ${}^{13}$C+${}^{13}$C by 30\% around $\emph{E}_{\rm cm}$=4.2 MeV. This difference has been explained by the coupling effect of neutron transfer in the ${}^{12}$C+${}^{13}$C fusion reaction \cite{esbensen2011}. At energies below 3 MeV, the effect from the transfer reaction becomes negligible and the two limitations agree with each other with a difference less than 17\%; a value similar to the experimental errors of the fusion cross section measurements at sub-barrier energies. 

The direct measurement by Spillane $\emph{et al.}$ \cite{spillane2007} reported a strong resonance at 2.14 MeV. Considering the large uncertainty, the direct measurement is only higher than the upper limit by 1.87$\sigma$. It has been discussed above that the indirect measurement does not support the existence of this resonance. Therefore one has to wait for a better measurement to tell whether or not such a resonance that exceeds the theoretical upper limit exists. 

The lower limits presented in the present work include an empirical lower limit (KNS) and a theoretical calculation (TWDP) \cite{zhang2020}. The current TDWP approach does not include the cluster effect and only provides a baseline for the ${}^{12}$C+${}^{12}$C S$^*$ factor at lower energies. It is interesting to note that the TDWP calculation agrees with the empirical lower limit (KNS) with a deviation less than 33\% at energies below 3 MeV. Combining the new upper limits with the empirical lower limit and the prediction of TDWP, the ${}^{12}$C+${}^{12}$C S$^*$ factors are better constrained in spite of the unknown resonances within the unmeasured energy range.

\section{The ${}^{12}$C+${}^{12}$C Reaction Rate}
\label{sec_reaction_rate}

\begin{figure}[!htbp]
\centering
\includegraphics[width=0.49\textwidth]{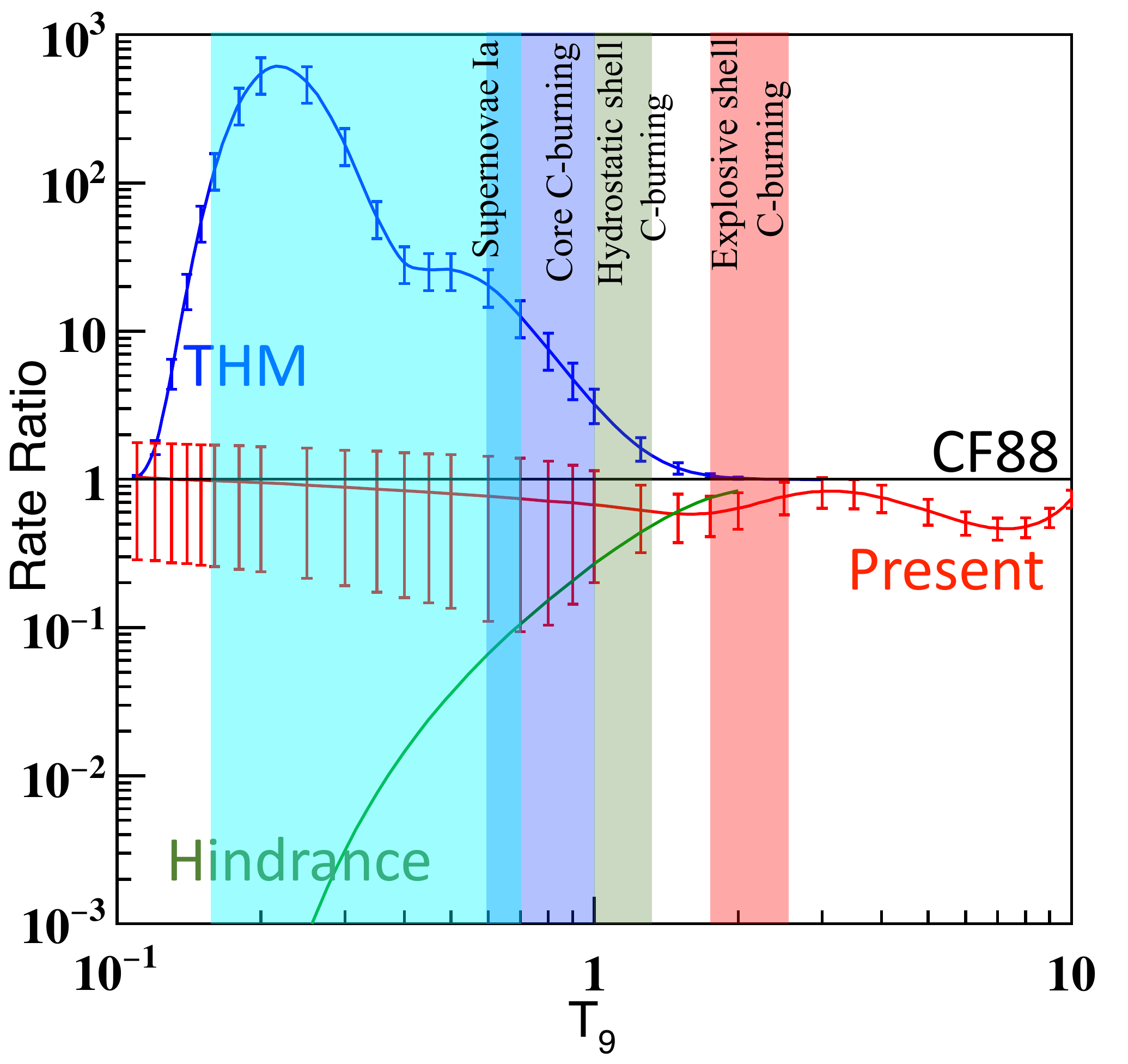}
\caption{The reaction rates relative to the CF88 rate \cite{caughlan1988} obtained in the present work (Red) together with the rates based on the THM measurement \cite{tumino2018} (Blue) and hindrance model \cite{esbensen2011} (Green). The temperatures for the type Ia supernovae (T=0.15-0.7 GK), the core carbon burning (T=0.6-1.0 GK), the hydrostatic shell carbon burning (T=1.0-1.2 GK), and the explosive shell carbon burning (T=1.8-2.5 GK) are marked by colored bands.}
\label{fig_rate_ratio}
\end{figure}

The reaction rate of ${}^{12}$C+${}^{12}$C is calculated with the measured S$^*$ factors at E$_{\rm cm}>$2.7 MeV, where a reasonable agreement among the experimental total S$^*$ factors exists, and using the lower and upper limits for $\emph{E}_{\rm cm}<$2.7 MeV. The corrected S$^*$ factor obtained with the Spillane $\emph{et al.}$ measurement is used in the range of 2.7 MeV to 4.4 MeV. The corrected S$^*$ factor obtained with the Kettner $\emph{et al.}$ measurement is used in the range of 4.4 MeV to 6.3 MeV. A $\pm$30\% uncertainty is used to account for the deviation among different experimental data sets as well as the systematic errors of the statistical model in addition to the statistical errors. The theoretical fusion cross section is used for higher energies \cite{hagino2015} with an assumed 10\% uncertainty to account for the experimental error. For the energies below 2.7 MeV, the upper limit obtained with ${}^{13}$C+${}^{13}$C and the lower limit from the TDWP prediction were used as they represent the highest and lowest limits, respectively. The assumption is made that there is no extremely strong resonance structure (such as the greatly enhanced rate values from 0.14 to 0.4 GK reported by Tumino $\emph{et al.}$ \cite{tumino2018}) for low-temperature rates. The average of these two limits is used as the averaged values for the S$^*$ factor. The resulting reaction rate is listed in Table \ref{tab_new_rate}. The ratio of the current reaction rate to the standard CF88 \cite{caughlan1988} rate is shown in Fig. \ref{fig_rate_ratio} together with the ones obtained with the hindrance model and the THM indirect measurement. 

\begin{table}[]
\caption{\small Recommended reaction rate for ${}^{12}$C+${}^{12}$C.}
\label{tab_new_rate}
\begin{tabular}{p{1.6cm} |p{2.3cm} |p{2.3cm} }
\hline
T$_9$                         & Rate ($cm^3 mol^{-1} s^{-1}$) & Relative Uncertainty (1$\sigma$) \\ \hline
0.11	&	7.61E-50	&	72\%	\\
0.12	&	1.06E-47	&	72\%	\\
0.13	&	8.78E-46	&	73\%	\\
0.14	&	4.70E-44	&	73\%	\\
0.15	&	1.75E-42	&	73\%	\\
0.16	&	4.76E-41	&	74\%	\\
0.18	&	1.65E-38	&	74\%	\\
0.2	&	2.53E-36	&	75\%	\\
0.25	&	5.97E-32	&	77\%	\\
0.3	&	1.27E-28	&	78\%	\\
0.35	&	5.73E-26	&	80\%	\\
0.4	&	8.77E-24	&	81\%	\\
0.45	&	6.13E-22	&	82\%	\\
0.5	&	2.36E-20	&	83\%	\\
0.6	&	9.48E-18	&	86\%	\\
0.7	&	1.12E-15	&	87\%	\\
0.8	&	5.64E-14	&	85\%	\\
0.9	&	1.53E-12	&	79\%	\\
1	&	2.60E-11	&	70\%	\\
1.25	&	7.17E-09	&	48\%	\\
1.5	&	5.07E-07	&	36\%	\\
1.75	&	1.56E-05	&	31\%	\\
2	&	2.76E-04	&	28\%	\\
2.5	&	2.56E-02	&	25\%	\\
3	&	7.18E-01	&	23\%	\\
3.5	&	8.98E+00	&	22\%	\\
4	&	6.47E+01	&	21\%	\\
5	&	1.17E+03	&	20\%	\\
6	&	8.87E+03	&	18\%	\\
7	&	3.99E+04	&	17\%	\\
8	&	1.27E+05	&	16\%	\\
9	&	3.20E+05	&	15\%	\\
10	&	6.79E+05	&	14\%	\\
\hline
\end{tabular}
\end{table}

\section{Discussion}

Both particle and $\gamma$-ray spectroscopies have been used to investigate the decay channels of ${}^{12}$C+${}^{12}$C. By summing the S$^*$ factors of the ground state transitions ($p_0$ and $\alpha_0$) and the transitions of main characteristic $\gamma$-rays (440 keV and 1634 keV) and correcting for the missing channels using a statistical model, an agreement within $\pm$30\% has been achieved among all the measurements at energies $\emph{E}_{\rm cm}>$4 MeV. The systematic uncertainty of the statistical model is estimated to be 9\%($1\sigma$). At lower energies, though the measurements of $\gamma$-ray spectroscopy \cite{spillane2007,kettner1977} agree with each other, there are still some discrepancies among the particle spectroscopy measurements. These disagreements arise from four possible sources: 1) underestimation of the beam induced background; 2) incomplete measurement of the decay channels populating the higher excited states or erroneous correction for the missing channels; 3) experimental errors, such as target thickness and effective energy determination; 4) the angular distribution of the particles and $\gamma$ rays. It has been pointed out that the assumption of isotropic angular distribution often used in the charged particle spectroscopy at lower energies may lead to a 20$\%$ error \cite{fang2017}. Complete measurements of the angular distributions with good statistics would be useful to reduce this error.

To suppress the background, the particle-$\gamma$ coincidence measurement was performed using the ATLAS accelerator at Argonne National Laboratory with coincidence between GammaSphere and large area strip silicon detectors \cite{jiang2018}. From the particle-$\gamma$ coincidence spectrum at $\emph{E}_{\rm cm}$=5.0 MeV from Ref. \cite{jiang2012}, the characteristic $\gamma$ rays in the p-channel can be clearly identified. The particle-$\gamma$ coincidence technique \cite{jiang2018} can eliminate background, and record the $p_i$-$\gamma$ and $\alpha_i$-$\gamma$ events with clean background. Of course, the random-coincidence events need to be well-studied to ensure accuracy. This measurement was further extended down to 2.16 MeV by the STELLA collaboration \cite{stella2018,fruet2020}. However a more recent measurement of the $\alpha_1$ channel \cite{tan2020} is nearly a factor of 10 less than the measurements of Becker$\emph{et al.}$ \cite{becker1981} and Jiang $\emph{et al.}$ \cite{jiang2018} around $\emph{E}_{\rm cm}$=2.94 MeV, calling for more measurements at these energies to resolve the discrepancies.

Fruet $\emph{et al.}$ \cite{fruet2020} and Tan $\emph{et al.}$ \cite{tan2020} both measured $p_1$-$\gamma$(440) and $\alpha_1$-$\gamma$(1634) coincidence events at very low energies, and then provided total cross sections for proton and alpha channels by normalizing their $\sigma_{p_1}$ and $\sigma_{\alpha_1}$ values using the ratios $\sigma_{p_1}$/$\sigma_{p}$, $\sigma_{\alpha_1}$/$\sigma_{\alpha}$ from Becker $\emph{et al.}$ \cite{becker1981}. In reference \cite{stella2018}, these ratios are $\sigma_{p_1}$/$\sigma_{p}$=(15.6$\pm$0.7)\%, $\sigma_{\alpha_1}$/$\sigma_{\alpha}$=(31.9$\pm$1.4)\%, which are only mean values, ignoring the huge fluctuation that exist (see the Fig. III.27 of the reference \cite{stella2018}). However, according to the present analysis in Fig. \ref{fig_ratio_p_channel} and Fig. \ref{fig_ratio_a_channel}, the standard deviations for the $\sigma_{p_1}$/$\sigma_{p}$ and $\sigma_{\alpha_1}$/$\sigma_{\alpha}$ ratios of Becker $\emph{et al.}$ \cite{becker1981} are 42\% and 26\%, respectively. Therefore, it is important to recognize that large uncertainties exist in the total cross sections of $\sigma_{p}$ and $\sigma_{\alpha}$ that are derived only from $\sigma_{p_1}$ and $\sigma_{\alpha_1}$ and their corresponding branching ratios $\sigma_{p_1}$/$\sigma_{p}$ and $\sigma_{\alpha_1}$/$\sigma_{\alpha}$ as was done in the recent Fruet $\emph{et al.}$ \cite{fruet2020} and Tan $\emph{et al.}$ \cite{tan2020} results.
 
Our statistical model calculation based on the data obtained with particle spectroscopy shows that the relative systematic uncertainties of the predicted branching ratios get smaller as the branching ratios increase. Although the particle-$\gamma$ coincidence measurements offer a clean background, deriving the total S$^*$ factors for the proton and alpha channels only from the $p_1$-$\gamma$(440) or $\alpha_1$-$\gamma$(1634) events at astrophysical energies would suffer from fluctuations arising from the complicated resonances in the $^{12}$C+$^{12}$C system shown in Fig. \ref{fig_ratio_p_channel} and Fig. \ref{fig_ratio_a_channel}. Furthermore the resonances of $p_1$ or $\alpha_1$ do not necessarily represent the resonances of p- or $\alpha$- channels. Therefore, in order to effectively derive the total fusion cross sections of $^{12}$C+$^{12}$C, we propose measuring the ground states of the $\alpha$- and p- channels ($\alpha_0$, $p_0$) together with the particle channels in coincidence with the 440, 1634, 2391, 2640 and 2982 keV characteristic $\gamma$ rays to cover the $p_1$, $\alpha_1$, $p_2$, $p_3$ and $p_4$ channels which would limit the systematic errors in the branching ratios. Our calculation also shows that a constant branching ratio is the best choice for deriving the S* factors of ${}^{12}$C(${}^{12}$C,$\alpha$)${}^{20}$Ne and ${}^{12}$C(${}^{12}$C,$p$)${}^{23}$Na from the measured partial S* factors.

The ground state transitions are crucial as these channels contribute significantly at stellar energies (Fig. \ref{fig_ratio_p_channel} and Fig. \ref{fig_ratio_a_channel}). New technologies are needed to suppress the backgrounds. As a complement for the particle-$\gamma$ coincidence technique, a solenoid spectrometer has been applied towards investigating the ground state decay channels of $^{12}$C+$^{12}$C fusion \cite{fang2017ssnap}. An efficient thick target method based on particle spectroscopy has shown great promise by scanning for the existence of potential resonances through a wide energy range with only a single incident energy \cite{tang2019}. Particle identification techniques such as $\Delta$E-E and TOF-E are useful to identify protons and $\alpha$ particles at energies around 1 MeV and effectively suppress the beam induced background and cosmic background \cite{zickefoose2018,stella2018}. The rather large contribution of the $\alpha$ transfer channel, $^{12}$C(${}^{12}$C,${}^{8}$Be)${}^{16}$O reported at energies below 3 MeV should be investigated both theoretically and experimentally to clarify its role at stellar energies.

The large uncertainty in the astrophysical reaction rate of ${}^{12}$C+${}^{12}$C is created mainly by the extremely low-trend predicted by the hindrance model \cite{esbensen2011}, which is independent of the resonance structure, and now the recent THM result \cite{tumino2018} which predicts a very high reaction rate (see Fig. \ref{fig_rate_ratio}). The resonance structure itself makes it difficult experimentally to discern between the various predictions. Our upper and lower limits are established empirical based on the carefully evaluated data obtained by direct measurements. These limits present a new way to extrapolate the ${}^{12}$C+${}^{12}$C S$^*$ factors down to the important energy region. They significantly reduce the existing uncertainties in the extrapolation. It is important to test these limits with the precise direct measurements to be performed at energies below 2.7 MeV. If there were currently unknown relatively strong resonances, which are fundamentally different from those already confirmed resonances observed by direct measurements at energies above 2.7 MeV, the upper and lower limits presented in the present work should be revised.

\section{Summary}

In summary, we calculated the branching ratios of many states in the $\alpha$- and p- channels for ${}^{12}$C+${}^{12}$C fusion based on the statistical model. The theoretical branching ratios are compared with the experimental branching ratios measured with particle spectroscopy. The theoretical results in the present work reproduce well the averaged trend of the branching ratios for $p_i$ and typical characteristic $\gamma$ rays (\emph{E}$_\gamma$= 440 keV) in ${}^{12}$C(${}^{12}$C,p)${}^{23}$Na, and $\alpha_i$ and typical characteristic $\gamma$ rays (\emph{E}$_\gamma$=1634 keV) in ${}^{12}$C(${}^{12}$C,$\alpha$)${}^{20}$Ne. Our calculations show that the relative systematic uncertainties of the predicted branching ratios get smaller as the predicted ratios increase. The various data sets obtained with $\gamma$ or particle spectroscopies are corrected for the missing channels and reasonable agreements among the S$^*$ factors of the proton, $\alpha$ and their summation are achieved at energies above 2.7 MeV. Upper and lower limits are recommended for energies below 2.7 MeV. A new ${}^{12}$C+${}^{12}$C reaction rate is also recommended. In addition, we also find that the indirect measurement done with the THM is found to be inconsistent with the direct measurement by comparing the thick target yield of the 1634 keV $\gamma$-ray transition. Better measurements at energies below 4 MeV are needed to resolve the discrepancies among the existing measurements. Reliable measurements at 2.7 MeV are particularly needed to guide the development of extrapolating models, calibrate the theory for indirect measurements and verify the proposed upper and lower limits used in the extrapolation. 

\begin{acknowledgments}
X.D. Tang acknowledges Dr. Kettner for providing the original data of his work. This work was supported by the National Key Research and Development Program of China under Grant No. 2016YFA0400501, the National Natural Science Foundation of China under Grant Nos. 11805291, 11575292, 11475228, 11490564, 11875329, the U.S. Department of Energy under Grant No. DE-AC07-05ID14517, the Fundamental Research Funds for the Central Universities under Grant No. 18lgpy84, the Continuous Basic Scientific Research Project under Grant No. WDJC-2019-13. X. D. Tang thanks the support of the Strategic Priority Research Program of Chinese Academy of Sciences (No. XDB34000000) and the ``Hundred Talents Program'' of the Chinese Academy of Sciences.
\end{acknowledgments}




\vspace{5mm}

\clearpage

\end{document}